\documentclass[pra,twocolumn,showpacs]{revtex4-1}
  
\usepackage{amsmath}
\usepackage{nicefrac}
\usepackage{graphicx}
\usepackage{microtype}
\usepackage{placeins}
\usepackage{physics}
\usepackage{mdframed}
\usepackage{mathrsfs}
\usepackage{nicefrac}
\usepackage{multirow}
\usepackage{placeins}

\usepackage{soul}
\usepackage[normalem]{ulem}

\usepackage[english]{babel}
\usepackage[applemac]{inputenc}
\usepackage[T1]{fontenc}

\newcommand{\beq}{\begin{equation}}
\newcommand{\eeq}{\end{equation}}
\newcommand{\beqa}{\begin{eqnarray}}
\newcommand{\eeqa}{\end{eqnarray}}

\renewcommand\Re{\operatorname{Re}}
\renewcommand\Im{\operatorname{Im}}

\usepackage{enumitem}
\setlist[description]{leftmargin=*}

\begin{document}

\title{Diagrammatic approach to orbital quantum impurities\\[3pt] interacting with a many-particle environment}
\author{G. Bighin and M. Lemeshko}
\date{\today}

\affiliation{IST Austria (Institute of Science and Technology Austria), Am Campus 1, 3400 Klosterneuburg, Austria}

\begin{abstract}

Recently it was shown that an impurity exchanging orbital angular momentum with a surrounding bath can be described in terms of the angulon quasiparticle [Phys. Rev. Lett. \textbf{118}, 095301 (2017)]. The angulon consists of a quantum rotor dressed by a many-particle field of boson excitations, and can be formed out of, for example, a molecule or a nonspherical atom in superfluid helium, or out of an electron coupled to lattice phonons or a Bose condensate. Here we develop an approach to the angulon based on the path-integral formalism, which sets the ground for a systematic, perturbative treatment of the angulon problem. The resulting perturbation series can be interpreted in terms of Feynman diagrams, from which, in turn, one can derive a set of diagrammatic rules. These rules extend the machinery of the graphical theory of angular momentum -- well known from theoretical atomic spectroscopy -- to the case where an environment with an infinite number of degrees of freedom is present. In particular, we show that each diagram can be interpreted as a `skeleton,' which enforces angular momentum conservation, dressed by an additional many-body contribution. This connection between the angulon theory and the graphical theory of angular momentum is particularly important as it allows to systematically and substantially simplify the analytical representation of each diagram. In order to exemplify the technique, we calculate the $1-$ and $2-$loop contributions to the angulon self-energy, the spectral function, and the quasiparticle weight. The diagrammatic theory we develop  paves the way to investigate next-to-leading order quantities in a more compact way compared to the variational approaches. 

\end{abstract}

\maketitle


\section{Introduction}

Impurity problems, where an isolated quantum particle interacts with a many-body environment, represent one of the key concepts in condensed matter, atomic, and chemical physics~\cite{Kondo:1964nea,Anderson:1967zze,Crawford:1972,Fukuhara:2013hq,PolaronsExcitons,EminPolarons,LeggettRMP87,WeissBook,Su:1980tx,Holstein:1959hg,Devreese:2009hz}. Studying quantum impurities amounts to an active, interdisciplinary research field of its own, with an additional motivation coming from the domain of strongly-correlated phases of matter. A quantum impurity, in fact, can be regarded as a building block for strongly-correlated systems, providing a basis to advance their understanding \cite{Massignan:2014gq,Alexandrov:1986,Zhao:1997,Alexandrov:1995}, as well as to develop more accurate numerical techniques \cite{Georges:1996zz}. 

Most impurities studied during last years are  structureless -- as in the case of an electron in a polarizable medium forming a polaron \cite{Landau:1948vr,Frohlich:1950fc,Frohlich:1954ds} -- or can be considered structureless due to a very large gap between the ground state and the first excited state. An example of the latter are polarons formed of an  atomic impurity immersed in an ultracold Bose or Fermi gas~\cite{ChikkaturPRL00, SchirotzekPRL09, PalzerPRL09, Tempere:2009gm, KohstallNature12, KoschorreckNature12, SpethmannPRL12, FukuharaNatPhys13, ScellePRL13, Cetina15, Massignan:2014gq, Jorgensen2016, Hu:2016in,  Cetina2016}. Another kind of well-studied impurity problems involves a localized spin coupled to a bath of fermions~\cite{LutchynPRB08}, bosons~\cite{LeggettRMP87},  or other spins~\cite{ProkofievSpinBath00}.

In several settings, however, an impurity possesses more involved degrees of freedom, such as orbital angular momentum.
For instance, transfer of orbital angular momentum from an electron to the phonon bath  is believed to provide a major contribution to ultrafast demagnetization of ferromagnetic thin films~\cite{StammPRB10, TsatsoulisPRB16, FahnleJSNM17}. On the other hand, molecular rotation is  known to be altered by the presence of a quantum solvent, such as superfluid $^4$He~\cite{Toennies:2004gc}. Furthermore, recent breakthroughs in the manipulation of ultracold quantum gases opened up the possibility to prepare ultracold diatomic molecules in  selected quantum rotational states and fine-tune the long-range interactions between them~\cite{Quemener:2012kq,Dulieu:2009ei,Carr:2009ch,Wall:2015, Ospelkaus:2010iq,Ospelkaus:2010cl,Takekoshi:2012cj,Takekoshi:2014ku, Park:2015dj,Will:2016ih,Grobner:2016go,Lemeshko:2013kk,YanNature13}.   This paves the way to study interactions between molecular impurities and the surrounding Bose or Fermi gas.

Recently, it was shown that interaction of such orbital impurities with a many-particle environment can be rationalized by using the concept of the angulon quasiparticle~\cite{Schmidt:2015hc, Schmidt:2016du, Lemeshko:2016, Lemeshko:2016ti, Yakaboylu16, Redchenko:2016dq, Li:2016vy}. While in the case of polarons the bath degrees of freedom couple to the impurity's translational motion, in angulons the orbital angular momentum is redistributed between the impurity and the many-particle environment. Quantum rotations, in turn, are described by the non-Abelian $SO(3)$ algebra and feature a discrete spectrum of eigenvalues~\cite{Varshalovich:1988}. As a result, the angulon   problem becomes substantially more involved and distinct from other impurity problems involving, e.g., the polaron~\cite{Devreese13} or spin-boson~\cite{LeggettRMP87, WeissBook} type of coupling.

The concept of angulons has been used to study  a variety of physical systems, ranging from molecular ions rotating in a  BEC \cite{Midya:2016un} to molecules in superfluid helium nanodroplets \cite{Lemeshko:2016ti, Redchenko:2016dq}, using variational approaches in either the strong- \cite{Schmidt:2016du, Li:2016vy} or weak-coupling \cite{Schmidt:2015hc, Yakaboylu16, Redchenko:2016dq} regimes. A strong evidence was provided that molecules rotating in superfluid $^4$He form angulon quasiparticles~\cite{Lemeshko:2016ti, YuliaPhysics17}.

The coupling of rotations to a bath has been extensively studied in the context of molecules in helium nanodroplets using density functional calculations~\cite{Hernando:2007fe}, a combination of semi-analytical and Monte Carlo techniques~\cite{Marx:1999cc,Zillich:2004haa}, reptation quantum Monte Carlo~{\cite{Skrbic:2007bg,Moroni:2004ji,Tang:2004gh,Moroni:2003cz}}, path integral Monte Carlo~{\cite{Marx:2006hb,RodriguezCantano:2016ci,Zillich:2007bf,Topic:2006fd,Paesani:2005jr,Kwon:2000hy,Kwon:1996jg}}, and diffusion Monte Carlo~{\cite{Viel:2007ey,vonHaeften:2006ct,Paolini:2005df,Paesani:2004bx,Zillich:2004cta,Paesani:2001dc,Kwon:2000hy,Lee:1999id,Blume:1996eg}}. All these techniques, however, model the environment as a cluster of a finite size, and -- as a consequence -- are computationally expensive. The angulon theory, on the other hand, accounts for an infinite number of degrees of freedom in the many-body environment  analytically, and leads to a computationally inexpensive description in terms of quasiparticles. In the present paper we develop a path integral and diagrammatic approaches to the angulon which allow to integrate out the many-body degrees of freedom exactly and thereby simplify the treatment further. Starting from the seminal papers by Feynman~\cite{Feynman:1955zz,Feynman:1962zz}, path integrals have constituted one of the sharpest theoretical tools available to study the Fr\"ohlich polaron, providing a superior, all-coupling treatment~\cite{Devreese:1993wo}.

The paper is organized as follows. In Section \ref{sec:piangulon} we describe the angulon quasiparticle using the path integral formalism. Here, the path integral serves two aims: on one hand it allows for an exact treatment of the many-body bosonic environment, leading to an effective, single-particle description of the angulon in terms of a quantum rotor with time-non-local self-interaction. On the other hand, we show that the path-integral description we develop, naturally leads to a diagrammatic expansion, derived in Section \ref{sec:perturbativeexpansion}. The diagrammatic expansion, in turn, can be carried out in the angular momentum basis systematically, leading to a peculiar set of Feynman rules mapping diagrams to  corresponding analytical expressions -- as shown in Section~\ref{sec:rules}. In order to illustrate the capabilities of the technique,  in Section~\ref{sec:dysoneq} we derive the Dyson equation in the angular momentum representation. Next, in Section~\ref{sec:sigmaaz} we calculate the $1-$loop and $2-$loop contributions to the angulon self-energy, to the spectral function, and to the quasiparticle weight.

The Feynman rules we obtain bear a remarkable resemblance to the rules one encounters in the context of the graphical theory of angular momentum, widely used in atomic and nuclear theory~\cite{Balcar:2009,Judd:1963,Rudzikas:2007, Varshalovich:1988}. In such a way, it becomes possible to establish a connection between atomic structure calculations  -- dealing with a finite number of particles --  and the many-particle physics featured by the angulon. In particular, we show that  each diagram can be decomposed into a `skeleton' -- which coincides with the corresponding diagram from the graphical theory of angular momentum -- dressed by an additional contribution accounting for the many-body character of the problem. This paves the way to employ the mathematical machinery developed the context of the graphical theory of angular momentum as a building block of many-body calculations involving an infinite number of interacting particles.

The framework we introduce provides a fast way of calculating higher order quantities -- corresponding to multiphonon processes -- which in the case of variational treatments \cite{Schmidt:2015hc} would require very involved angular momentum algebra, making use of $3nj$ symbols for a $n$-phonon process.

\section{Path integral description of the angulon}
\label{sec:piangulon}

The starting point is the angulon Hamiltonian \cite{Schmidt:2015hc,Schmidt:2016du,Lemeshko:2016}, describing an orbital impurity exchanging angular momentum with a many-body environment:
\beq
\hat{H} = \hat{H}_\text{imp} + \hat{H}_\text{bos} + \hat{H}_\text{imp-bos},
\label{eq:h}
\eeq
where $\hat{H}_\text{imp}$ and $\hat{H}_\text{bos}$ give the kinetic energies of the impurity and the bosonic bath, respectively, and $\hat{H}_\text{imp-bos}$ describes the impurity-bath interactions. As mentioned above, the formalism can be used to describe a variety of the systems, from highly-excited electronic states~\cite{Balewski:2013hv} and cold molecules~\cite{Midya:2016un}  interacting with a BEC, to electrons exchanging orbital angular momentum with a crystal lattice~\cite{Stamm:2007hy}, to polyatomic species   embedded in superfluid helium nanodroplets~\cite{Lemeshko:2016ti}. For the sake of concreteness,  we will think of the impurity as of a linear rotor molecule, as described by the following Hamiltonian:
\beq
\hat{H}_\text{imp} = B \hat{\mathbf{J}}^2,
\eeq
where the rotational constant $B=1/(2 I)$ is expressed through the molecular moment of inertia,  $I$, and the units of $\hbar \equiv 1$ are used hereafter. The bosonic environment is described by the second term in Eq. (\ref{eq:h}), namely
\beq
\hat{H}_\text{bos} = \sum_{k \lambda \mu} \omega_{k} \hat{b}^\dagger_{k \lambda \mu} \hat{b}_{k \lambda \mu}, \\
\eeq
where $\sum_k \equiv \int \mathrm{d} k$, $\omega_k$ is the dispersion relation for bath excitations and the $\hat{b}^\dagger$ ($\hat{b}$) operator creates (destroys) a bosonic excitation with linear momentum $k$, angular momentum, $\lambda$, and angular momentum projection along the $z$ axis, $\mu$. The field operators in the angular momentum basis are defined in terms of the usual field operators as
\beq
b^\dagger_{k \lambda \mu} = \frac{k}{(2 \pi)^{\nicefrac{3}{2}}} \int \mathrm{d} \Omega_\mathbf{k} b^\dagger_{\mathbf{k}} \mathrm{i}^{-\lambda} Y_{\lambda \mu} (\Omega_\mathbf{k}),
\label{eq:bklm}
\eeq
and similarly for $b_{k \lambda \mu}$, having introduced the spherical harmonics $Y_{\lambda \mu}$~\cite{Varshalovich:1988} and the spherical coordinate representation of the vector $\mathbf{k}$, namely $\mathbf{k} \to \left\{ k, \Omega_\mathbf{k}\right\}$, with $\Omega_\mathbf{k}=\left\{ \theta_\mathbf{k}, \phi_\mathbf{k} \right\}$, see Ref.~\onlinecite{Lemeshko:2016} for details. Finally, the interaction between the molecule and the bosonic environment is given by the following term:
\beq
\hat{H}_\text{imp-bos} = \sum_{k \lambda \mu} U_\lambda(k) \left[ Y^*_{\lambda \mu} (\hat{\theta},\hat{\phi}) \hat{b}^\dagger_{k \lambda \mu} + Y_{\lambda \mu} (\hat{\theta},\hat{\phi}) \hat{b}_{k \lambda \mu} \right],
\label{eq:hmolbos}
\eeq
where $U_\lambda(k)$ is the angular-momentum-dependent potential in momentum-space, and the operators $(\hat{\theta}, \hat{\phi})$ give the orientation of the molecular impurity with respect to the laboratory frame. Here, only two Euler angles are required in order to describe a linear molecule. In the most general case (such as that of symmetric and asymmetric top molecules) the interaction~\eqref{eq:hmolbos} will depend upon the third Euler angle, $\hat{\gamma}$.

In order to proceed with a path-integral description of the angulon, it is necessary to rewrite the Hamiltonian $\hat{H}$ in terms of the position and momentum operators~\cite{Feynman:1955zz,Feynman:1972,Schulman:1996}, as given by the following relations:
\begin{align}
\label{qpb1}
\hat{q}_{k \lambda \mu} = \sqrt{\frac{1}{2 m \omega_k}} \left( \hat{b}_{k \lambda \mu} + (-1)^\mu \ \hat{b}^\dagger_{k \lambda -\mu} \right) \\
\label{qpb2}
\hat{p}_{k \lambda \mu} = - \mathrm{i} \sqrt{\frac{m \omega_k}{2}} \left( \hat{b}_{k \lambda \mu} - (-1)^\mu \ \hat{b}^\dagger_{k \lambda -\mu}\right)
\end{align}
This definition is analogous to the usual expressions of the ladder operators for the standard harmonic oscillator, with $m$ being the mass of each particle constituting the bosonic environment, and the angular momentum basis operators are related to the usual momentum-space operators in a complete analogy to Eq. (\ref{eq:bklm}).  After the substitutions of \eqref{qpb1} and \eqref{qpb2}, the Hamiltonian~\eqref{eq:h} reads
\beq
\hat{H} = B \hat{\mathbf{J}}^2 + \sum_{k \lambda \mu} \frac{1}{2m} |\hat{p}_{k \lambda \mu}|^2 + \frac{m \omega^2_k}{2} |\hat{q}_{k \lambda \mu}|^2 + \gamma_{k \lambda \mu} \hat{q}_{k \lambda \mu}
\label{eq:hbos3}
\eeq
where we have introduced
\beq
\gamma_{k \lambda \mu} (\hat{\theta},\hat{\phi}) = \sqrt{2 m \omega_k} U_\lambda (k) Y_{\lambda \mu}(\hat{\theta},\hat{\phi}) \; .
\eeq
By Legendre-transforming the Hamiltonian of Eq.~(\ref{eq:hbos3}), we obtain the corresponding Lagrangian. Next, integrating over time and replacing each field operator with a corresponding field variable, we arrive at the action
\begin{multline}
S[q(t),\Omega(t)] = \int \mathrm{d} t B \mathbf{J}^2 +\\+ \int \mathrm{d} t \sum_{k \lambda \mu} \frac{m}{2} |\dot{q}_{k \lambda \mu}|^2 - \frac{m \omega^2_k}{2} |q_{k \lambda \mu}|^2 - \gamma_{k \lambda \mu} (\theta, \phi)q_{k \lambda \mu}
\label{eq:s0}
\end{multline}
Here it is implied that  $\mathbf{J}$ is a differential operator acting on the rotor coordinates, $\Omega(t)$. With the action at hand, we can reformulate the angulon problem in terms of path integral. Let us consider the Green function describing the total amplitude for a particle to evolve in time from the configuration $\Omega_i = \left\{ \theta_i, \phi_i \right\}$ to the configuration $\Omega_f = \left\{ \theta_f, \phi_f \right\}$ during time $T$. The invariance of the theory under time translations ensures that the Green function is a function of time differences only. Within the path-integral formalism, it can be written as a sum over all possibile trajectories connecting $\Omega_i$ and $\Omega_f$, weighted by a factor $\exp(\mathrm{i}S)$:
\beq
G(\Omega_i,\Omega_f; T)=\int_{\substack{\Omega(0)=\Omega_i\\ \Omega(T)=\Omega_f}} \mathcal{D} \Omega \prod_{k \lambda \mu} \mathcal{D} q_{k \lambda \mu} \ e^{\mathrm{i} S[q(t),\Omega(t)]}
\eeq

The first part of the integration measure, $\mathcal{D} \Omega$, corresponds to the rotating molecule, while the second part, $ \prod_{k \lambda \mu} \mathcal{D} q_{k \lambda \mu}$, describes the many-body environment. Crucially, the integration over $q_{k \lambda \mu}$ can be carried out exactly as the $q$ field appears quadratically and linearly in the action \cite{Grosche:1998,Khandekar:1993,Kleinert:2009}, leading to the following result:
\begin{equation}
G(\Omega_i,\Omega_f; T)=\int_{\substack{\Omega(0)=\Omega_i\\ \Omega(T)=\Omega_f}} \mathcal{D} \Omega \ e^{\mathrm{i} S_\text{eff}[\Omega(t)]}
\label{eq:g0a}
\end{equation}
(the boundary conditions for the path integral will be omitted from now on). The effective action reads

\beq
\begin{split}
S_\text{eff} = &\underbrace{\int_0^T \mathrm{d} t B \mathbf{J}^2}_{S_0} + \\ &+ \underbrace{\frac{\mathrm{i}}{2} \int_0^T \mathrm{d} t \int_0^T \mathrm{d} s \sum_\lambda P_\lambda (\cos \gamma (t,s)) \mathcal{M}_{\lambda} (|t-s|)}_{S_\text{int}}
\end{split}
\label{eq:effectives}
\eeq
Here $P_\lambda$ are the Legendre polynomials, $\gamma(t,s)$ is the angle between the position of the rotor at time $t$ and at time $s$, and $\mathcal{M}$ is defined as
\beq
\mathcal{M}_\lambda(|t-s|) = \frac{2 \lambda + 1}{4 \pi} \sum_{k} |U_\lambda (k)|^2 \ e^{- \mathrm{i} \omega_k |t-s|}
\eeq
Equation (\ref{eq:effectives}) is the main result of the present section: the first term, $S_0$, describes a free linear rotor, whereas the second term, $S_\text{int}$, accounts for the interaction of the rotor with its past self. Thus, analogously to the path-integral treatment of the Fr\"ohlich polaron \cite{Feynman:1955zz,Feynman:1962zz,Tempere:2009gm}, the bath degrees of freedom can be integrated out exactly, leading to an effective single-particle description, in which an effective potential encodes the many-body physics of the original problem. In contrast to the polaron, however,  the orbital impurity considered here is moving in the internal space represented by the non-abelian $SO(3)$ group, rather than in the usual three-dimensional space. This  makes the angulon problem substantially different and more involved compared to the polaron problem~{\cite{Lemeshko:2016}}.

\section{Diagrammatic expansion}
\label{sec:perturbativeexpansion}

In order to investigate the properties of the angulon through the effective action of Eq. (\ref{eq:effectives}), we pursue a perturbative expansion, also dubbed as direct path-integral treatment in the context of polarons~\cite{Kholodenko:1983iu,vonBaltz:1972td,Novikov:2010cx}. Starting from the definition of the angulon's Green function, Eq.~(\ref{eq:g0a}), we treat the interaction term $S_\text{int}$ as a perturbation. Then, the perturbation series for the angulon Green function can  be written as
\begin{align}
G (\Omega_i, \Omega_f ; T) &= G_0 (\Omega_i, \Omega_f ; T) + \sum_{n=1}^{\infty} \frac{\mathrm{i}^n}{n!} \langle S_\text{int}^n \rangle_0
\label{eq:exp0}
\end{align}
Here $\langle X \rangle_0 \equiv \int \mathcal{D} \Omega \ X \exp(\mathrm{i} S_0)$ denotes the expectation value of $X$ taken over the states of the free impurity, as described by $S_0$ alone, and
\beq
G_0(\Omega_i,\Omega_f; T )= - \mathrm{i} \sum_{\lambda \mu} Y_{\lambda \mu}(\Omega_i) Y_{\lambda \mu}^*(\Omega_f) e^{-\mathrm{i} B \lambda (\lambda + 1) T}
\label{eq:g0begin}
\eeq
is the Green function of a free linear rotor~\cite{Favro:1960ie},  also see Appendix \ref{app:gf}. We note that the order of magnitude of the perturbation parameter $S_\text{int}$ is determined by the potential term $|U_\lambda (k)|^2$, making the present perturbation theory essentially a weak-coupling theory, as it will be confirmed later by a comparison with other angulon theories. Analyzing the perturbation series, one notices that the $0$th order term coincides with the free propagator $G_0$, whereas the $1$st order term reads 
\beq
G^{(1)} (\Omega_i, \Omega_f ; T) = - \frac{\mathrm{i}}{2}  \int \mathcal{D} \Omega \ e^{\mathrm{i} S_0} \int \mathrm{d} t \mathrm{d} s \ \chi (t,s)
\label{eq:g1}
\eeq
with the shorthand
\beq
\chi(t,s) = - \mathrm{i} \sum_\lambda P_\lambda (\cos \gamma(t,s)) \mathcal{M}_\lambda(|t-s|) \;.
\eeq
By introducing two midpoints in the path integral at times $t$ and $s$, and integrating over the angular configurations at the midpoints, one can rewrite Eq. (\ref{eq:g1})   in terms of the propagators $G_0$ and $\chi$. For shortness' sake we introduce the new variables, $i$, $f$, $1$, and $2$, bundling together the angular configuration and time, e.g. $1=\left\{ \Omega_1, t_1 \right\}$, so that the first order contribution reads
\beq
G^{(1)} (i,f) = - \frac{\mathrm{i}}{2} \int \mathrm{d} 1 \mathrm{d} 2 \ G_0 (i, 1) \ G_0 (1,2) \ \chi (1,2) \ G_0 (2,f)
\label{eq:g1if}
\eeq
Eq.~\eqref{eq:g1if} has a simple interpretation in terms of Feynman diagrams:

\begin{center}
\includegraphics[width=.40\linewidth]{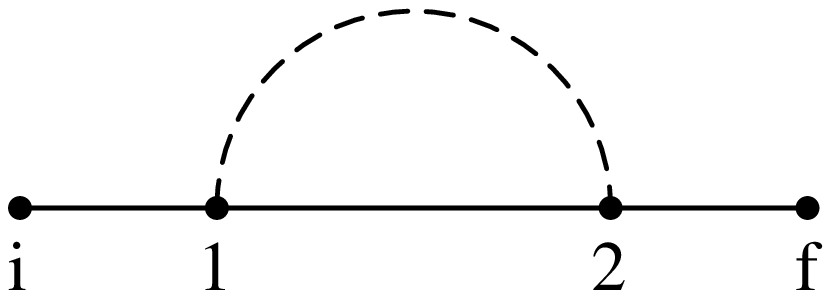}
\end{center}
where a solid line represents the free propagator $G_0$ and a dashed line corresponds to $\chi$.  The meaning $\chi$ thereby becomes clear: it is the phonon propagator, describing the interaction of the angulon with the many-body environment. The same reasoning can be straightforwardly generalized to the $n$th order contribution to the angulon Green function. By making use of $2n$ midpoints, we obtain:
\begin{widetext}
\beq
G^{(n)} (i,f) = \frac{1}{2^n} \frac{(-\mathrm{i})^n}{n!} \sum_{\{p_i\}} \int \mathrm{d} 1 \ldots \mathrm{d} 2n \ G_0 (i,1) \ldots G_0 (2n,f) \chi ({p_1},{p_2}) \ldots \chi ({p_{2n-1}},{p_{2n}})
\label{eq:gn}
\eeq
\end{widetext}
where the sum extends over all the permutations, $\{ {p_1}, \ldots, {p_{2n}} \}$, of the space-time configurations, $\{ 1, \ldots, {2n} \}$. These permutations give rise to various topologically distinct Feynman diagrams. For instance, at the second order we get the following set of diagrams from $G^{(2)}(i,f)$:
\begin{center}
\includegraphics[width=.60\linewidth]{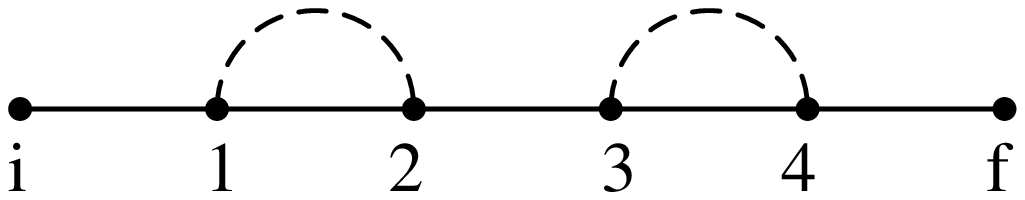} \\
\includegraphics[width=.60\linewidth]{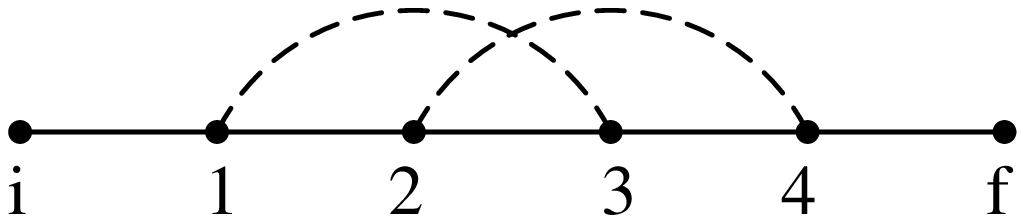} \\
\includegraphics[width=.60\linewidth]{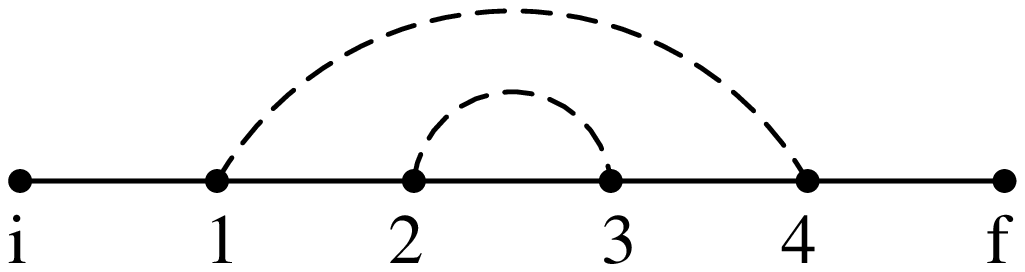}
\end{center}

Finally, let us discuss the combinatorial prefactor in Eq. (\ref{eq:gn}), following the argument presented in Ref. \cite{Kholodenko:1983dy}. Let us start by considering a single interaction line, $\chi(a,b)$, connecting two configurations at times $t_a$ and $t_b$. Clearly, as both time variables run from $0$ to $T$, one may have either $t_a > t_b$ or $t_b > t_a$. Alternatively and equivalently, we can choose to use the retarded propagator for the interaction $\chi$, which is non-zero only for $t_b>t_a$, and multiply the final result by a factor of $2$ to account for the original multiplicity. This reasoning yields a factor of $2^n$ when applied to $n$ interaction lines at $n$th order. Having fixed the time ordering for every interaction line coupling two configurations, we can still choose the relative time ordering of configurations not connected by an interaction line. This can be achieved by using a retarded propagator also for $G_0$ and thereby enforcing the `natural' time ordering for all configurations, i.e. $t_1 > t_2 > \ldots > t_{2n}$, selecting one possible ordering among $n!$ combinations, and therefore requiring another prefactor $n!$. Thus we have demonstrated that, when enforcing the `natural' time ordering by means of retarded propagators, every term in the perturbative series has no combinatorial prefactor \cite{Kholodenko:1983dy,Kholodenko:1983iu} as the prefactor $1/(2^n n!)$ in Eq. (\ref{eq:gn}) cancels out. In what follows, we will always use this convention, introducing the natural time ordering for the time variables, making use of retarded propagators, and omitting the combinatorial prefactors.

\section{Feynman rules for the angulon}
\label{sec:rules}

\begin{table*}[ht!]
\caption{\label{table1}Feynman rules for the angulon in the angular momentum basis. The prescription for the sign of each $\mu$ is given in the  text.}
\begin{tabular}{|c|c|}
\hline
\hline
Each external line & \multirow{2}{*}{$\sum_{\lambda_i \mu_i} (-1)^{\mu_i} G_{0,\lambda_i} \delta_{\lambda_\text{ext},\lambda_i} \delta_{\mu_\text{ext},\pm \mu_i}$} \\
\includegraphics[width=.15\linewidth]{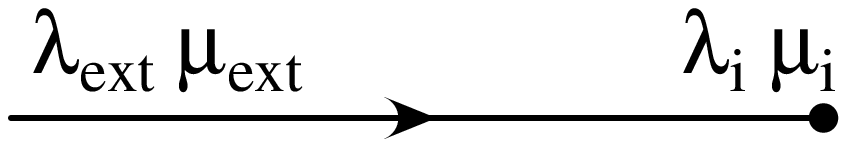}  & \\
\hline
Each internal $G_0$ line & \multirow{2}{*}{$ \sum_{\lambda_i \mu_i} (-1)^{\mu_i} G_{0,\lambda_i}$} \\
\includegraphics[width=.15\linewidth]{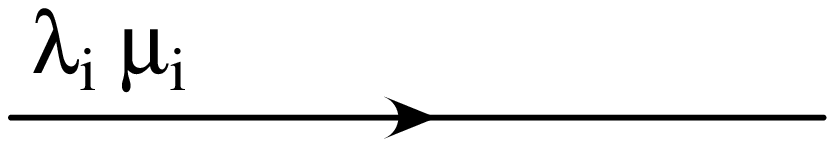}  & \\
\hline
Each internal $\chi$ line & \multirow{2}{*}{$ \sum_{\lambda_i \mu_i} (-1)^{\mu_i} \chi_{\lambda_i}$} \\
\includegraphics[width=.15\linewidth]{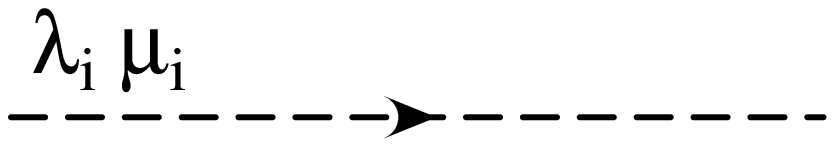}  & \\
\hline
Each vertex & \multirow{2}{*}[-16pt]{$ (-1)^{\lambda_i} \bra{\lambda_i} |Y^{(\lambda_j)}| \ket{\lambda_k} \begin{pmatrix}
\lambda_i & \lambda_j & \lambda_k \\
\mu_i & \mu_j & \mu_k
\end{pmatrix} $} \\
\includegraphics[width=.12\linewidth]{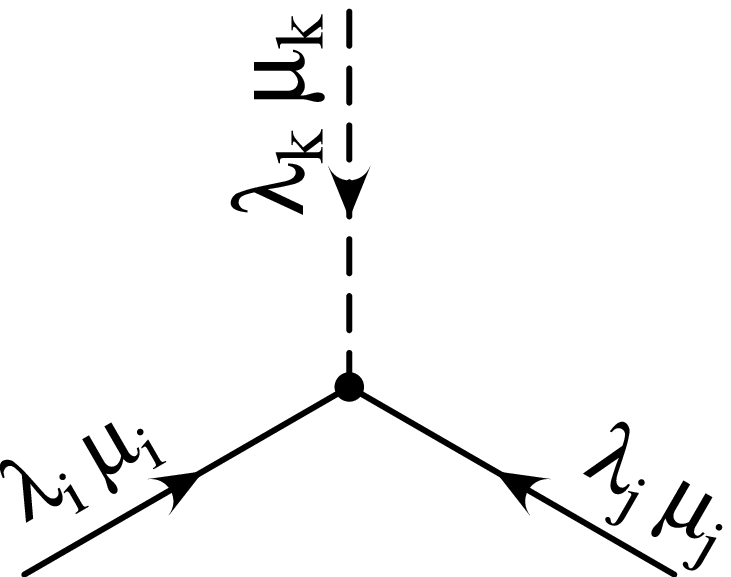}  & \\
\hline
\hline
\end{tabular}
\end{table*}

The aim of the present section is to establish a link between the diagrams corresponding to every term in the perturbative series generated by Eq. (\ref{eq:gn}) and their analytic expressions written in the angular momentum basis, by analogy with the usual Feynman rules in momentum space \cite{Popov:2001,Veltman:1994}. The motivation for switching to momentum-space diagrams comes from the great simplification of the analytic expressions we can achieve. As one can see, direct evaluation of a single $n$th order term of   Eq.~({\ref{eq:gn}})  requires a calculation of  a $4n$-dimensional integral over the angles. The diagrams in the angular momentum space, on the other hand, require the evaluation of a discrete sum over the angular momentum, $\lambda$, and its projection along the $z$ axis, $\mu$, for each internal line of the corresponding diagram, in addition to $n$ frequency integrations. Furthermore, we will show that the frequency integrations, as well as the sums over $\mu$, can be carried out analytically at every order. When working with structureless particles, the aforementioned simplification arises naturally in the momentum basis, as a consequence of the translational invariance of Green functions~{\cite{Kleinert:2001}}. Here, the angular momentum basis is the right choice, as a consequence of the rotational invariance of each Green function.

We now outline how these new rules arise in the angular momentum representation, considering the angular and time dependence of each Green function separately.

\subsubsection{Angular dependence}

Let us take into account a diagram representing a contribution to the angulon Green function. For consistency, let us consider  $G^{(1)} (\Omega_i, \Omega_f)$ of Eq. (\ref{eq:g1if}), however the same reasoning can be applied at every order. We introduce the expansion in the angular momentum basis for a function of two angular variables, defined as
\beq
G^{(1)}_{\lambda \mu l m} = \int \mathrm{d} \Omega_i \mathrm{d} \Omega_f \ Y^*_{\lambda \mu}(\Omega_i) Y_{lm}(\Omega_f) \ G^{(1)} (\Omega_i, \Omega_f)
\label{eq:g1a}
\eeq
In order to proceed, we need to express every quantity inside the integral   in the angular momentum basis. Hence, we replace each propagator $G_0$ and $\chi$, included in Eq.~(\ref{eq:g1a}) through Eq.~(\ref{eq:g1if}), with its representation in the angular momentum basis, defined as
\beq
G_0(\Omega,\Omega') = \sum_{\lambda \mu l m} Y_{\lambda \mu}(\Omega) Y^*_{lm}(\Omega') \ G_{0, \lambda \mu l m}
\label{eq:gzo}
\eeq
and
\beq
\chi (\Omega,\Omega') = \sum_{\lambda \mu l m} Y_{\lambda \mu}(\Omega) Y^*_{lm}(\Omega') \ \chi_{\lambda \mu l m} \; .
\label{eq:gzo2}
\eeq
The representation in Eqs.~(\ref{eq:gzo}) and (\ref{eq:gzo2}) can be greatly simplified due to rotational invariance. Using the lemma introduced in Appendix \ref{app:lemma2} we can rewrite them as
\beq
G_0(\Omega,\Omega') = \sum_{\lambda \mu} Y_{\lambda \mu} (\Omega) Y^*_{\lambda \mu} (\Omega') G_{0,\lambda}
\label{eq:g0s}
\eeq
and
\beq
\chi(\Omega,\Omega') = \sum_{\lambda \mu} Y_{\lambda \mu} (\Omega) Y^*_{\lambda \mu} (\Omega') \chi_{\lambda} \; .
\label{eq:g0schi}
\eeq
The angulon Green function in the angular momentum basis, $G_{0,\lambda}$, and the interaction Green function, $\chi_\lambda$, are calculated in Appendix \ref{app:gf}.

After inserting the momentum-space representations of each Green function appearing in Eq. (\ref{eq:g1a}), the algebra gets quite involved. Nonetheless, after some manipulations, a clear pattern -- valid for every diagram and at every order in the diagrammatic expansion -- emerges, and the angular momentum representation of each diagram follows the prescriptions listed below.

\begingroup
\setlength{\parindent}{0cm}

\begin{description}

\item[Lines] Each line in a diagram corresponds to a set of quantum numbers $\{ \lambda_i, \mu_i\}$, representing the angular momentum and its projection along the $z$ axis. Each line must be oriented in an arbitrary, however consistent way -- graphically we draw an arrow on each line. We have introduced these labels in Figures~\ref{fig:first} and~\ref{fig:second}, which illustrate the one- and two-loop contributions to the angulon self-energy studied in the next section. When transcribing a line, one needs to write the propagator in the angular momentum representation:  either a free propagator,  $(-1)^{\mu_i} G_{0,\lambda_i}$, for a solid line, or an interaction  propagator, $(-1)^{\mu_i} \chi_{\lambda_i}$, for a dashed line~\footnote{The $(-1)^{\mu_i}$ factors before each propagator are obtained by using the following property of the spherical harmonics $Y^{*}_{\lambda \mu} (\theta, \phi) = (-1)^{\mu} Y_{\lambda -\mu} (\theta, \phi)$.}. In addition, for every line one needs to write a summation over the corresponding quantum numbers, i.e. $\sum_{\lambda_i, \mu_i}$. An additional integration over the phonon momentum $k$ for interaction lines is contained in $\chi_\lambda$, see Eq. (\ref{eq:chifinalnoapp}).

\item[External lines] The integrations over $\Omega_i$ and $\Omega_f$ from Eq. (\ref{eq:g1a}), i.e. the integrations over the configurations which belong to an external line in a Feynman diagram,  give a result of the form
\beq
\int \mathrm{d} \Omega_i Y^{*}_{\lambda_\text{ext}, \mu_\text{ext}} (\Omega_i) Y_{\lambda_i, \pm \mu_i} (\Omega_i) = \delta_{\lambda_\text{ext}, \lambda_i} \delta_{\mu_\text{ext}, \pm \mu_i}
\label{eq:externall}
\eeq
where $\lambda_\text{ext}, \mu_\text{ext}$ are the quantum numbers associated with the external (initial or final) state. The sign of $\mu_i$ is determined by the orientation of the line, and is given by a $+$ ($-$) when the line is entering (leaving) the diagram. The resulting Kronecker $\delta$'s can be used to carry out the sums over the quantum numbers corresponding to external lines, removing the summation over the corresponding quantum numbers $\{ \lambda_i, \mu_i \}$.

\item[Vertices] Each vertex corresponds to an integral over three spherical harmonics, leading to~\cite{Varshalovich:1988}
\beq
V^{\lambda_i \lambda_j \lambda_k}_{\pm \mu_i \pm \mu_j \pm \mu_k} = (-1)^{\lambda_i} \bra{\lambda_i} |Y^{(\lambda_j)}| \ket{\lambda_k} \begin{pmatrix}
\lambda_i & \lambda_j & \lambda_k \\
\pm \mu_i & \pm \mu_j & \pm \mu_k
\end{pmatrix}
\label{eq:3jsymbols}
\eeq
where $\{\lambda_i,\mu_i\}$, $\{\lambda_j,\mu_j\}$ and $\{\lambda_k,\mu_k\}$ are the quantum numbers associated with each of the three lines entering the vertex, which are to be read in the counterclockwise direction. In Eq.~\eqref{eq:3jsymbols} we have introduced the $3j$ symbol and the reduced matrix element of the spherical harmonic operator, $\bra{\lambda_i} |Y^{(\lambda_j)}| \ket{\lambda_k}$~\cite{Varshalovich:1988} . Again, the signs of each $\mu_i$ are + ($-$) when the corresponding line is entering (leaving) the vertex. We note that the reduced matrix element in each vertex reflects the dynamics of the problem, whereas the $3j$ symbol encodes the information about the geometry. This point will be addressed in detail when analysing the structure of two-loop diagrams in Section~\ref{sec:sigmaaz}.

\end{description}
\endgroup

\subsubsection{Time dependence} 

The time dependence of the Green function follows the usual Feynman rules in momentum space~\cite{Veltman:1994,Kleinert:2001}. When taking the Fourier transform of the time-dependence of a diagram, we observe that each internal loop corresponds to an integral of the type $(2 \pi)^{-1} \int \mathrm{d} \omega_i$, each internal/external leg corresponds to a Fourier-transformed propagator,
\beq
G_{0,\lambda} (\omega) = \frac{1}{\omega - B \lambda (\lambda + 1) + \mathrm{i} \delta},
\label{eq:g0endnoapp}
\eeq
or
\beq
\chi_\lambda (\omega) = \sum_k \frac{|U_\lambda(k)|^2}{\omega - \omega_k + \mathrm{i \delta}},
\label{eq:chifinalnoapp}
\eeq
as derived in Appendix \ref{app:gf}, and the energy conservation throughout the whole diagram is enforced by choosing an adequate labelling, such as in Figures~\ref{fig:first} and~\ref{fig:second}.

\subsubsection{Discussion}

The approach just outlined can be systematically extended to every order in perturbation theory and leads to the rules listed in Table \ref{table1}, allowing us to bypass the lengthy expression of Eq. (\ref{eq:gn}). In order to evaluate a quantity at order $n$, one has to write all the relevant Feynman diagrams, and convert them to integrals in angular momentum space using the Feynman rules.
 
As opposed to most diagrammatic expansions~\cite{Abrikosov:1965,Altland:2006,Kholodenko:1983iu}, here the momentum integrals are replaced by discrete sums of $\lambda$ and $\mu$, which can be calculated exactly in a majority of cases or approximated numerically using a cutoff $\lambda_\text{max}$ to a very high precision~\cite{StoneBook13, SzalewiczIRPC08}.  Finally, we stress that the rules of Table \ref{table1} bear a remarkable resemblance with the rules derived in the context of the graphical theory of angular momentum~\cite{Balcar:2009,Judd:1963,Rudzikas:2007}, specifically the $3j$ symbols enforcing the angular momentum conservation at every vertex and, as well as the sign convention for the $\mu$ indices. In the present context, however, each line is `dressed' with a novel $G_0$ or $\chi$ propagator, reflecting the many-body character of the angulon. This connection with the graphical theory of angular momentum will be made clear in the Section \ref{sec:sigmaaz}, where we establish a rigorous mapping between the two theories and  use the graphical theory to simplify the angular momentum algebra.

\section{The Dyson equation}
\label{sec:dysoneq}

\begin{figure}[bh]
\centering
    \includegraphics[width=.75\linewidth]{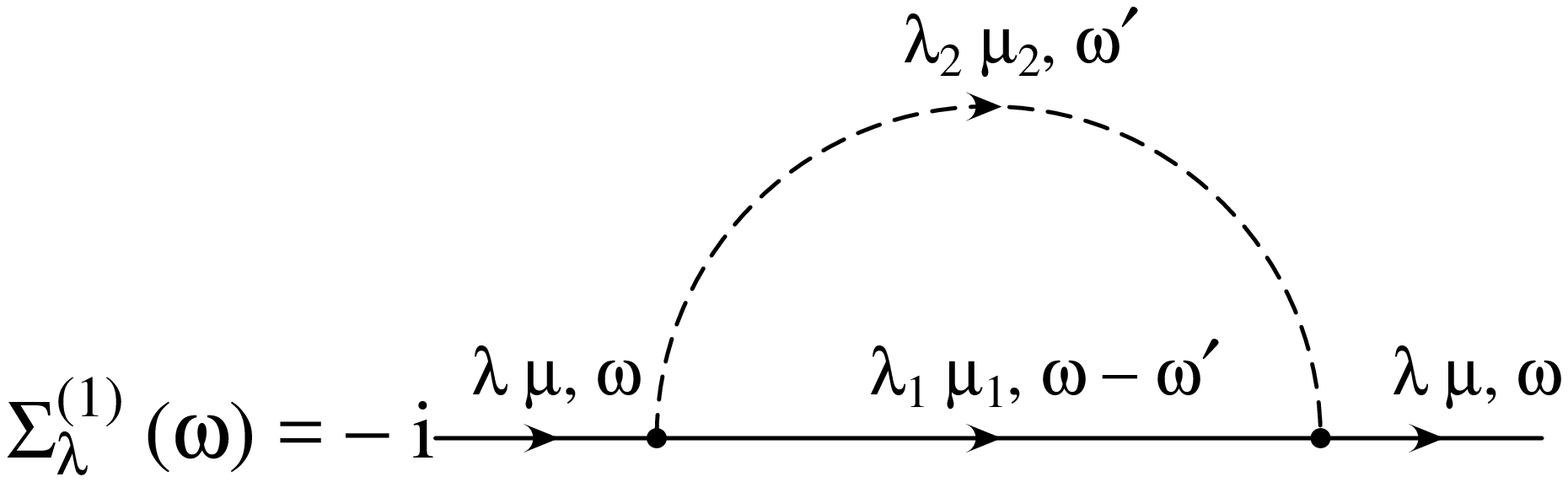}
\caption{The only diagram contributing to the first-order self-energy in the diagrammatic expansion. The labels $\lambda$ and $\omega$ on each solid (dashed) line denote the angular momentum and the energy of the angulon (phonon), respectively.}
\label{fig:first}
\end{figure}

The central object in the study of quasiparticles is the self-energy, $\Sigma$, which encompasses the renormalization of the quasiparticle properties due to the interaction with the many-body environment \cite{Abrikosov:1965}. Within the diagrammatic expansion, $\Sigma$ is identified as the $1-$particle-irreducible (1PI) contribution to the Green function. Here, it  corresponds to all the diagrams generated by  Eq. (\ref{eq:gn}) which cannot be divided into two by cutting a single internal line, with the external legs $G_0(i,1)$ and $G_0(2n,f)$ removed. The first-order contribution to the Green function of Fig. \ref{fig:first} is therefore 1PI, as well as the first and the second diagrams in Fig. \ref{fig:second}, which correspond to the second-order contribution. The third second-order diagram, however, can be divided into two by cutting a single internal line, thereby being reducible. With these definitions, the Green function of Eq. (\ref{eq:gn}) can be readily defined as an infinite series with alternating free propagators and the self-energy contribution. This infinite series can, in turn, be rewritten in a compact form as the Dyson equation for $G$ \cite{Dyson:1949ha}

\beq
G(i,f) = G_{0}(i,f) + \int \mathrm{d} 1 \mathrm{d} 2 \ G_{0}(i,1) \Sigma(1,2) G(2,f)
\label{eq:dyson}
\eeq
When working with structureless particles, Eq.~(\ref{eq:dyson}) greatly simplifies when rewritten in the frequency-momentum representation, due to the convolution theorem transforming each integral over the position in space and  time into a product. In the present case, the internal degrees of freedom of the angulon are represented by the angular configuration, $\Omega$, and the Fourier transform is replaced by the spherical harmonics expansion, as introduced in Eq. (\ref{eq:g1a}). Crucially, in Appendix~\ref{app:convolution} we demonstrate that  the rotational analogue of the convolution theorem holds in the angular momentum basis, allowing us to write the Dyson equation for the angulon as follows:
\beq
G_{\lambda} (\omega) = G_{0,\lambda} (\omega) + \sum_{n=1}^\infty (\Sigma_{\lambda} (\omega) G_{0,\lambda} (\omega))^n
\eeq
Summing the geometric series, we finally obtain a closed expression for the angulon Green function
\beq
G_{\lambda} (\omega) = \frac{1}{G_{0,\lambda}^{-1} (\omega) - \Sigma_\lambda (\omega)} \; .
\label{eq:gff}
\eeq
Clearly, the self-energy $\Sigma_\lambda (\omega)$ -- containing the 1PI contributions to the Green function at every order -- cannot be calculated in closed form. Nevertheless, the present formalism allows for a simple calculation of the first and second order terms (and, potentially, at  higher orders), as it will be demonstrated in the following section.

\section{Self-energy, spectral function, and quasiparticle weight}
\label{sec:sigmaaz}

\subsection{Self-energy}

Using the rules derived in Section \ref{sec:rules}, we associate the following analytic expression to the first-order self-energy diagram:
\begin{multline}
\Sigma^{(1)}_\lambda (\omega) = (-\mathrm{i}) \sum_{\lambda_1, \mu_1, \lambda_2, \mu_2} (-1)^{\mu_1+\mu_2} V^{\lambda,\lambda_1,\lambda_2}_{\mu,-\mu_1,-\mu_2} V^{\lambda,\lambda_1,\lambda_2}_{-\mu,\mu_1,\mu_2}  \times \\
\times \int \frac{\mathrm{d} \omega'}{2 \pi} G_{0,\lambda_1} (\omega - \omega') \chi_{\lambda_2} (\omega')
\label{eq:sigma1a}
\end{multline}
The integral over $\mathrm{d} \omega'$ in Eq. (\ref{eq:sigma1a}) can be evaluated exactly using contour integration in the complex plane. Moreover, using the properties of the $3j$ symbol~\cite{Varshalovich:1988} we can carry out the sums over $\mu_1$ and $\mu_2$, bringing Eq. (\ref{eq:sigma1a}) to the following form:
\beq
\Sigma^{(1)}_\lambda (\omega) = \frac{1}{4 \pi} \sum_{l_1, l_2,k} \frac{(2 l_1 + 1) (2 l_2 + 1) |U_{l_2} (k)|^2}{\omega_k + B l_1 (l_1+1) - \omega} \begin{pmatrix}
\lambda & l_2 & l_1 \\
0 & 0 & 0
\end{pmatrix}^2
\label{eq:sigma1x}
\eeq
By rewriting the $3j$ symbol in terms of the Clebsch-Gordan coefficients~\cite{Varshalovich:1988} we immediately recover the result found in Refs.~\cite{Schmidt:2015hc,Lemeshko:2016} using a variational ansatz for the wavefunction based on a single bath excitation. This strongly suggests that the diagrammatic expansion for the self-energy is equivalent to an expansion of the many-body wavefunction in bath excitations. We note that an analogous result holds for a spin-$\downarrow$ impurity in a Fermi sea of spin-$\uparrow$ fermions, where the equivalence between the variational ansatz including single particle-hole excitations and a diagrammatic treatment has been demonstrated \cite{Combescot:2007kz}.

\begin{figure}[bh]
\centering
  \begin{tabular}{c}
    \vspace{8pt} \\
    \includegraphics[width=.85\linewidth]{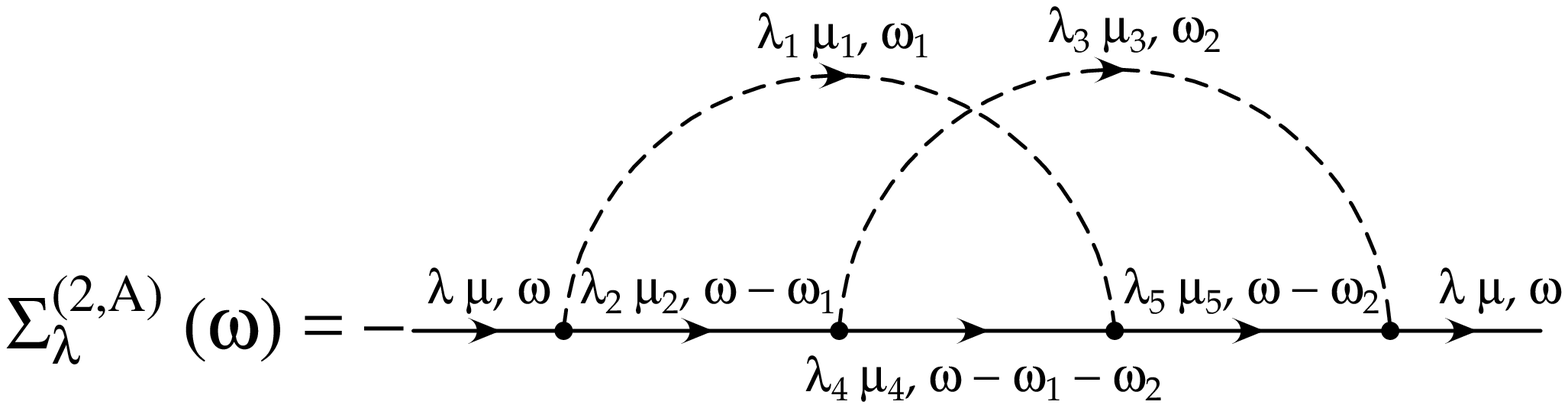} \\
    \vspace{8pt} \\
    \includegraphics[width=.85\linewidth]{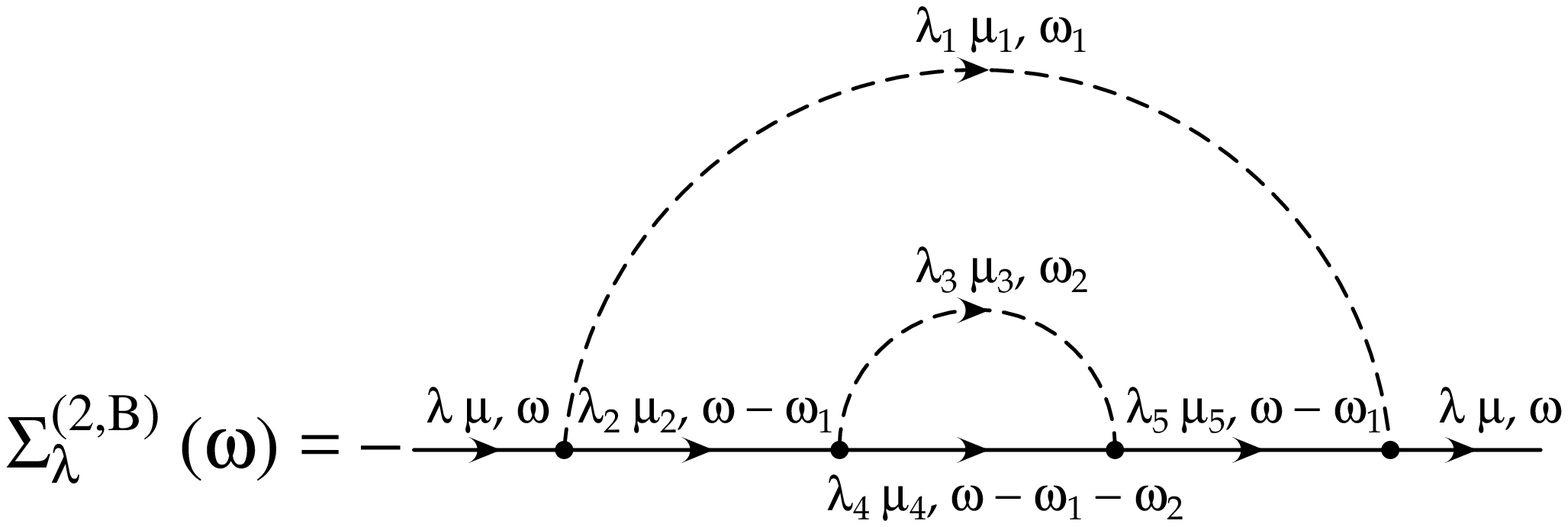} \\
    \vspace{8pt} \\
    \includegraphics[width=.85\linewidth]{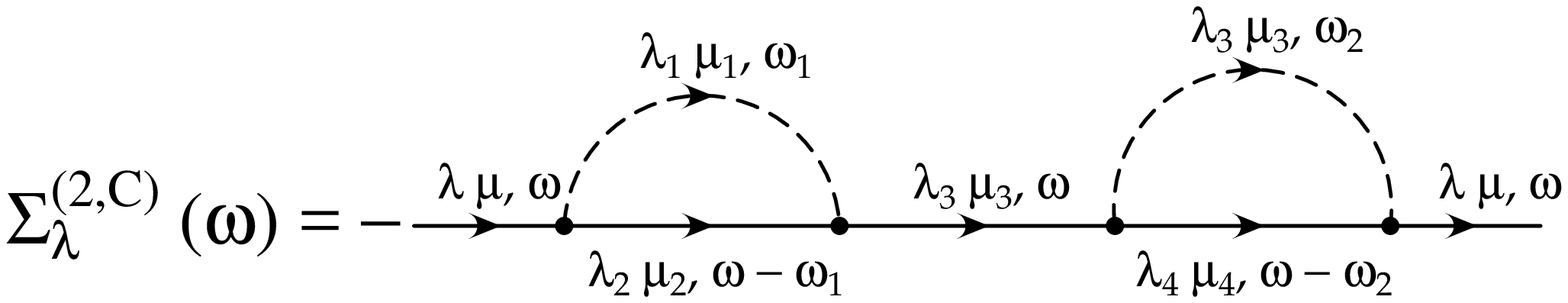}
  \end{tabular}
\caption{Diagrams appearing at the second order of the diagrammatic expansion. The first and second diagrams are $1-$particle irreducible and form the second order contribution to the self-energy $\Sigma^{(2)}$, whereas the third diagram is not $1-$particle irreducible and is accounted for in the Dyson sum for $\Sigma^{(1)}$.}
\label{fig:second}
\end{figure}

Now let us focus on the second-order contribution to the self-energy. Two topologically distinct contributions to the second-order self-energy -- as depicted in the first and second panel of Fig. \ref{fig:second} -- correspond to the following analytic expressions:
\begin{widetext}
\begin{multline}
\Sigma^{(2,A)}_\lambda (\omega) = (-\mathrm{i})^2 \sum_{\lambda_1,\mu_1, \ldots, \lambda_5,\mu_5} (-1)^{\sum_i \mu_i} V^{\lambda,\lambda_1,\lambda_2}_{\mu,-\mu_1,-\mu_2} V^{\lambda_2,\lambda_3,\lambda_4}_{\mu_2,-\mu_3,-\mu_4} V^{\lambda_1,\lambda_4,\lambda_5}_{\mu_1,\mu_4,-\mu_5} V^{\lambda_3,\lambda_5,\lambda}_{\mu_3,\mu_5,-\mu} \times \\ \times \int \frac{\mathrm{d} \omega_1}{2 \pi} \frac{\mathrm{d} \omega_2}{2 \pi} G_{0 \lambda_2} (\omega-\omega_1) G_{0 \lambda_4} (\omega-\omega_1-\omega_2) G_{0 \lambda_5} (\omega-\omega_2) \chi_{\lambda_1} (\omega_1) \chi_{\lambda_3} (\omega_2)
\label{eq:sigma2a}
\end{multline}
and
\begin{multline}
\Sigma^{(2,B)}_\lambda  (\omega) = (-\mathrm{i})^2 \sum_{\lambda_1,\mu_1, \ldots, \lambda_5,\mu_5} (-1)^{\sum_i \mu_i} V^{\lambda,\lambda_1,\lambda_2}_{\mu,-\mu_1,-\mu_2} V^{\lambda_2,\lambda_3,\lambda_4}_{\mu_2,-\mu_3,-\mu_4} V^{\lambda_3,\lambda_4,\lambda_5}_{\mu_3,\mu_4,-\mu_5} V^{\lambda_1,\lambda_5,\lambda}_{\mu_1,\mu_5,-\mu} \times \\ \times \int \frac{\mathrm{d} \omega_1}{2 \pi} \frac{\mathrm{d} \omega_2}{2 \pi} G_{0 \lambda_2} (\omega-\omega_1) G_{0 \lambda_4} (\omega-\omega_1-\omega_2) G_{0 \lambda_5} (\omega-\omega_1) \chi_{\lambda_1} (\omega_1) \chi_{\lambda_3} (\omega_2)
\label{eq:sigma2b}
\end{multline}
\end{widetext}
and
\beq
\Sigma_\lambda^{(2)} =\Sigma_\lambda^{(2,A)} + \Sigma_\lambda^{(2,B)} \; .
\eeq

Let us analyse the structure of the two-loop self-energies in detail, in order to reveal their physical meaning. It is convenient to split the analytic expressions corresponding to each self-energy diagram of Eqs. (\ref{eq:sigma2a}) and (\ref{eq:sigma2b}) into three terms, as

\beq
\Sigma_\lambda^{(2,x)} = (-\mathrm{i})^2 \ \sum_{\left\{ \lambda_i \right\}}\Sigma_\lambda^{(2,x) \text{dyn}} \times \Sigma_\lambda^{(2,x) \text{geom}} \times \Sigma_\lambda^{(2,x) \text{freq}}
\eeq
and $x=\left\{A,B\right\}$. The first term, $\Sigma_\lambda^{(2,x) \text{dyn}}$, contains the product of four reduced matrix elements for the spherical harmonic operator which describe the dynamics of angular momentum, namely
\begin{multline}
\Sigma_\lambda^{(2,A) \text{dyn}} =  \bra{\lambda} |Y^{(\lambda_1)}| \ket{\lambda_2}  \bra{\lambda_2} |Y^{(\lambda_3)}| \ket{\lambda_4} \times \\ \times  \bra{\lambda_1} |Y^{(\lambda_4)}| \ket{\lambda_5}  \bra{\lambda_3} |Y^{(\lambda_5)}| \ket{\lambda}
\end{multline}
and
\begin{multline}
\Sigma_\lambda^{(2,B) \text{dyn}} =  \bra{\lambda} |Y^{(\lambda_1)}| \ket{\lambda_2}  \bra{\lambda_2} |Y^{(\lambda_3)}| \ket{\lambda_4} \times \\ \times  \bra{\lambda_3} |Y^{(\lambda_4)}| \ket{\lambda_5}  \bra{\lambda_1} |Y^{(\lambda_5)}| \ket{\lambda} \; .
\end{multline}
The second term, $\Sigma_\lambda^{(2,x) \text{geom}}$,  containing a phase factor, the product of four $3j$ symbols and a summation over $\mu_i$, describes the geometric  aspects of the problem, i.e. the conservation of angular momentum. This term  can be understood   in terms of the graphical theory of angular momentum~\cite{Balcar:2009,Judd:1963,Rudzikas:2007}, in which analytical expressions involving angular momenta are rewritten as diagrams, often allowing for very substantial simplifications of lengthy calculations. Indeed, it turns out that when the `geometric' term is represented using the rules of the graphical theory of angular momentum, the resulting diagram has exactly the same topological structure as the diagrams of Fig.~\ref{fig:second}, provided that two external lines are joined and that every dashed interaction line is replaced by a solid line. This connection is analysed in greater detail in Appendix~\ref{app:graphical}, where it is shown that the summations over $\mu_i$ can be carried out exactly, leading to the following analytic expression
\beq
\Sigma_\lambda^{(2,A) \text{geom}} = \frac{(-1)^{\lambda_1 + \lambda_2 + \lambda_3 + \lambda_4}}{2 \lambda + 1} \begin{Bmatrix}
\lambda_2 & \lambda_1 & \lambda \\
\lambda_5 & \lambda_3 & \lambda_4
\end{Bmatrix}
\label{eq:s2ageo}
\eeq
having introduced the $6j$ symbol~\cite{Varshalovich:1988} and
\beq
\Sigma_\lambda^{(2,B) \text{geom}}
 = \frac{(-1)^{\lambda_4 + \lambda_5} \delta_{\lambda_2,\lambda_5}}{(2 \lambda + 1)(2 \lambda_5 + 1)} \{\lambda~\lambda_1~\lambda_2\} \{\lambda_2~\lambda_3~\lambda_4\}
\label{eq:s2bgeo}
\eeq
Here $\{ a~b~c \}$ is the $0j$ symbol~\cite{Sobelman:1995,Balcar:2009}, which equals $1$ if $a$, $b$, $c$ satisfy the triangular condition, and is $0$ otherwise.

Finally, the third term, $\Sigma_\lambda^{(2,x) \text{freq}}$, contains the frequency integrals and the summations over the phonon momenta. The former can be evaluated exactly using contour integration in the complex plane. Note that the two integrals, although very similar, are essentially different, reflecting different topological structure of the diagrams they represent. The results of contour integration are:
\begin{widetext}
\beq
\label{eq:sigma2afreq}
\Sigma_\lambda^{(2,A) \text{freq}} = \sum_{k_1, k_2} \frac{|U_{\lambda_1} (k_1)|^2 |U_{\lambda_3} (k_2)|^2}{(\omega-E_{\lambda_2} - \omega_{k_1}) (\omega-E_{\lambda_5} - \omega_{k_2}) (-\omega + E_{\lambda_4} + \omega_{k_1} + \omega_{k_2})}
\eeq
and
\beq
\label{eq:sigma2bfreq}
\Sigma_\lambda^{(2,B) \text{freq}} = \sum_{k_1, k_2} \frac{|U_{\lambda_1} (k_1)|^2 |U_{\lambda_3} (k_2)|^2}{(\omega-E_{\lambda_2} - \omega_{k_1}) (\omega-E_{\lambda_5} - \omega_{k_1}) (-\omega + E_{\lambda_4} + \omega_{k_1} + \omega_{k_2})}
\eeq
\end{widetext}
with $E_\lambda = B \lambda (\lambda+1)$. We stress that the integrals of Eq. (\ref{eq:sigma2afreq}) and Eq. (\ref{eq:sigma2bfreq}) -- along with the one-loop counterpart of Eq. (\ref{eq:sigma1x}) -- are the only equations in the present paper that need to be evaluated numerically in order to get the results in the present Section. Due to the moderate dimensionality of the integrals, the computation can be carried out with great accuracy using standard numerical libraries.

In conclusion of the present Section, let us comment on the connection to the graphical theory of angular momentum~{\cite{Balcar:2009,Judd:1963,Rudzikas:2007}}. Here, we have demonstrated that each diagram contains a `geometric' part which enforces angular momentum conservation. This part and can be understood -- and significantly simplified -- in terms of a completely analogous diagram introduced within the graphical theory of angular momentum. On the other hand, the `dynamical' and `frequency' parts associated with each diagram represent a novel contribution, and can be understood as a many-body part dressing the skeleton provided by the geometric terms.

\begin{figure*}[t]
\centering

    \includegraphics[width=.82\linewidth]{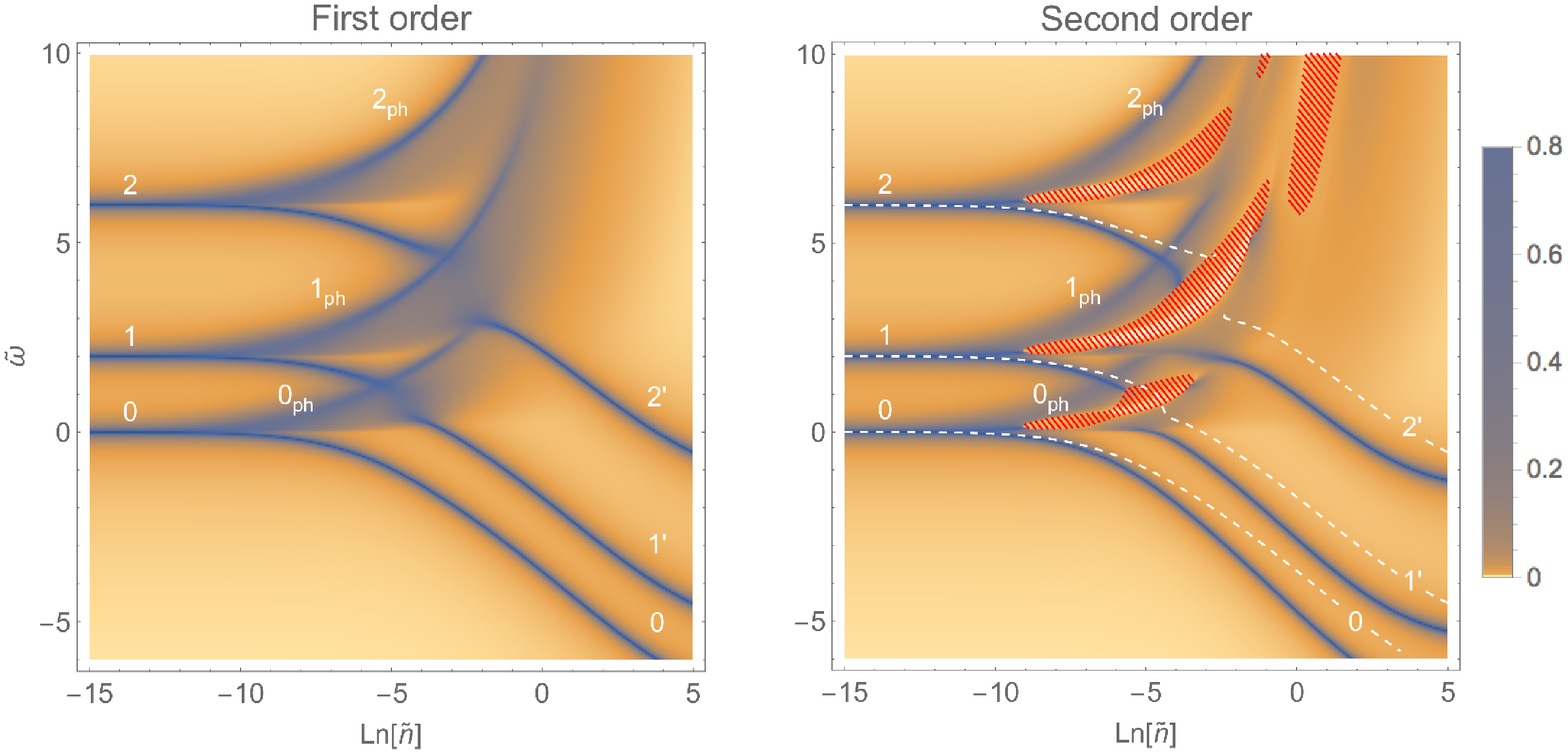}

\caption{The angulon spectral function $\tilde{\mathcal{A}}_L \equiv \mathcal{A}_L B$ for $L=0,1,2$ as a function of the dimensionless density, $\tilde{n} = n (m B)^{-\nicefrac{3}{2}}$, and of the dimensionless energy, $\tilde{\omega}=\omega/B$, parameters defined in the text. The left panel shows the first-order spectral function, Eq. (\ref{eq:spectralf}), obtained from the Dyson equation of Eq. (\ref{eq:gff}) and one-loop diagrams, Eq. (\ref{eq:sigma1x}). The right panel, on the other hand, includes two-loop contributions, Eqs. (\ref{eq:sigma2a}) and (\ref{eq:sigma2b}).  The white dashed lines show the energy of the first-order quasiparticle states, derived from Eq. (\ref{eq:pole}), showing the negative shift in the quasiparticle energy due to the inclusion of two-phonon processes in the high-density region. The dashed red region corresponds to an unphysical region with negative spectral weight, as described in the  text. The notation for the state labels is also introduced in the text.}
\label{fig:spectrals}
\end{figure*}

\subsection{Spectral function}
\label{subsec:spectralfunction}
The self-energy we have just calculated allows us to evaluate the angulon Green function through Eq.~(\ref{eq:gff}), which, in turn, leads to the angulon spectral function
\beq
\mathcal{A}_L (\omega) = - \frac{1}{\pi} \Im G_L (\omega + \mathrm{i} 0^+)
\label{eq:spectralf}
\eeq
The spectral function encodes the information about the angulon excitation spectrum, as well as its quasiparticle properties~\cite{Abrikosov:1965,Altland:2006,Schmidt:2015hc}.

\begin{figure}[t]
\centering

    \includegraphics[width=.96\linewidth]{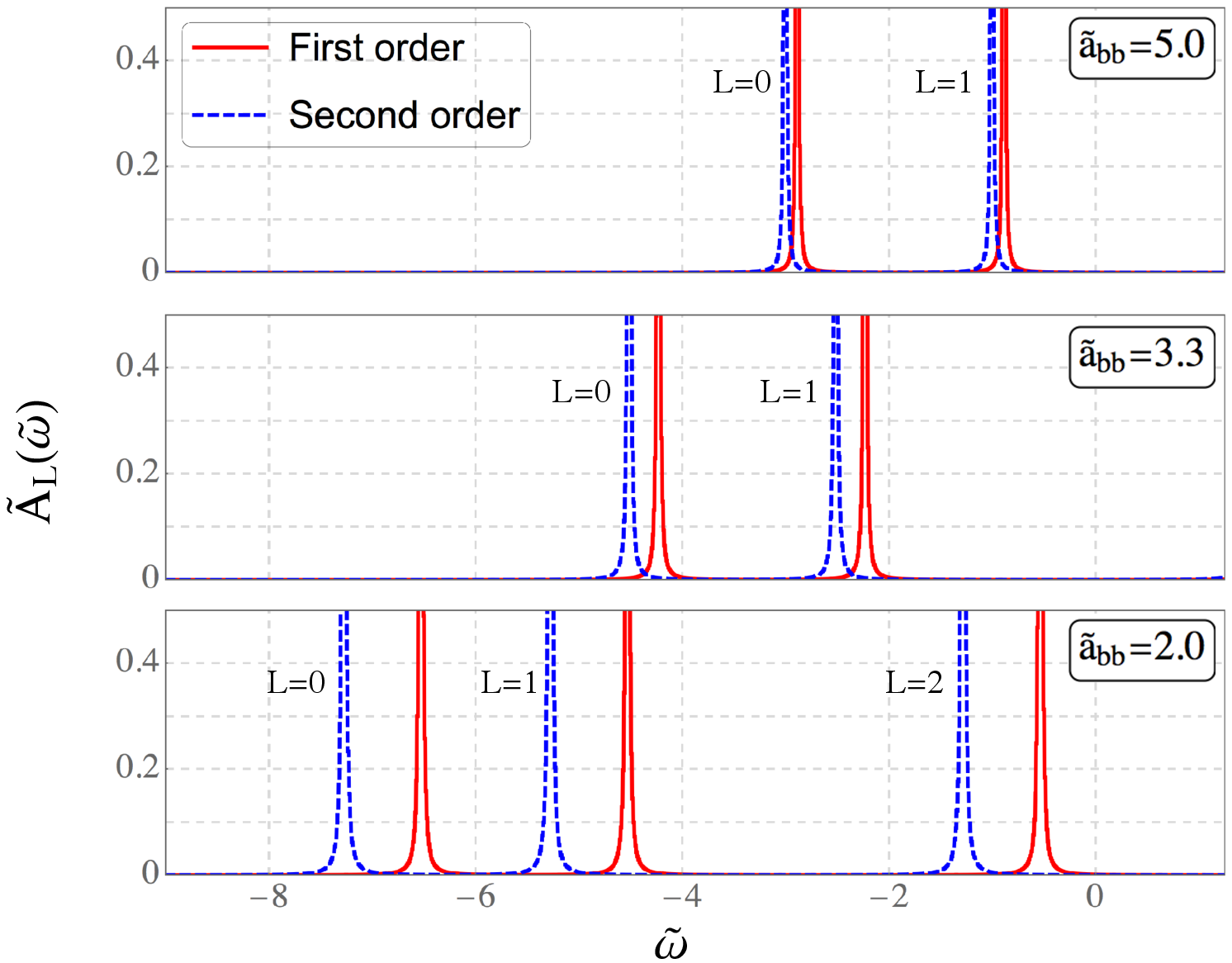}

\caption{The angulon spectral function, $\tilde{\mathcal{A}}_L \equiv \mathcal{A}_L B$, as a function of the dimensionless energy, $\tilde{\omega}=\omega/B$, in the high-density regime $\log(\tilde{n}) = 5$, for different values of the boson-boson scattering length (in units of $(m B)^{-\nicefrac{1}{2}}$). The lowest peaks for $L=0,1$ are shown, in the case of $a_{\text{bb}}=2.0 (m B)^{-\nicefrac{1}{2}}$ also a third peak $L=2$ is visible in the plotted range of energies. Smaller values of the boson-boson scattering length $a_{\text{bb}}$ correspond to a stronger renormalisation of the angulon energy, as well as to an increased splitting between the peaks calculated using the first order theory (solid line) and the second order theory (dashed lines).}
\label{fig:peaks}
\end{figure}

In order to analyze  the angulon spectral function quantitatively, let us define the   quantities introduced in Section~\ref{sec:piangulon}. We choose the effective momentum-space potential, $U_\lambda(k)$ of Eq. (\ref{eq:hmolbos}), to be of the same form as the one used in Ref. \onlinecite{Schmidt:2015hc} to describe an ultracold molecule immersed in a weakly-interacting BEC:
\beq
U_\lambda(k) = u_\lambda \sqrt{\frac{8 n k^2 \epsilon_k}{\omega_k (2 \lambda+1)}} \int \mathrm{d} r r^2 f_\lambda(r) j_\lambda(kr)
\eeq
where $n$ is the density of the bosonic bath, $\epsilon_k = k^2/(2m)$, $j_\lambda$ are the spherical Bessel functions and the form factors $f_\lambda$ determine the details of the molecule-environment interaction, along with the interaction parameters $u_\lambda$. We choose the same Gaussian form factors as in Ref.~\onlinecite{Schmidt:2015hc}, i.e. $f_\lambda(r)=(2 \pi)^{-\nicefrac{3}{2}} \exp(-r^2/(2 r_\lambda^2))$, as well as the same interaction parameters $u_0 = 1.75 u_1 = 218 B$. For the bosonic bath we take  the dispersion relation $\omega_k = \sqrt{\epsilon_k ( \epsilon_k + 2 g_\text{bb} n)}$, with $g_\text{bb}=4 \pi a_\text{bb}/m$. The boson-boson scattering length is set to $a_{\text{bb}}=2.0 (m B)^{-\nicefrac{1}{2}}$. Since the goal of this paper is to introduce a new formalism for the angulon, we use the same parameters as in Ref. \onlinecite{Schmidt:2015hc}, except for a reduced $a_\text{bb}$, in order to make the second-order corrections more evident. The role of $a_\text{bb}$ in enhancing the relevance of second-order corrections will also be analysed in Fig. \ref{fig:peaks}. We stress, however, that in the case of a molecular impurity within a He nanodroplet, the parameters of the model can also be inferred, in a more physical way, from the impurity-bath potential energy surfaces \cite{Lemeshko:2016ti, YuliaPhysics17}.

In Fig. \ref{fig:spectrals} we compare the dimensionless angulon spectral function $\tilde{\mathcal{A}}_L = \mathcal{A}_L B$ obtained using the one-loop self-energy (left panel) with the spectral function obtained from the one- and two-loop contributions (right panel), as a function of the dimensionless angulon energy $\tilde{\omega}=\omega/B$ and of the dimensionless density $\tilde{n} = n (m B)^{-\nicefrac{3}{2}}$. We briefly comment on the essential features of the spectral function, motivated by emphasizing the new features introduced by two-phonon processes analyzed for the first time in the present paper; a thorough description of the whole many-body-induced fine structure (MBIFS) can be found in Ref. \onlinecite{Schmidt:2015hc}. In the low density region the quasiparticle peaks essentially coincide with the energy levels of a free quantum rotor $E_L = B L (L + 1)$, so that we can simply label a state with its quantum number $L$.

As the density is increased the state splits and an upper phonon branch $L_\text{ph}$ develops, while the energy of the quasiparticle state is shifted towards lower energies. The $L=0$ state is stable across the whole parameter space considered, whereas the $L=1$ and $L=2$ states undergo an `angulon instability' for intermediate density values, corresponding to the emission of a phonon carrying a quantum of angular momentum, bringing the angulon to the $1'$ and $2'$ state, respectively.

Let us now focus on the modifications induced by the inclusion of two-phonon processes. Indeed, a comparison between the left and the right panel shows that the qualitative picture and the MBIFS is essentially unaltered, as no new features appear in the angulon spectral function. However, a closer look at the quantitative details reveals the relevance of two-phonon processes. We compare the position of the quasiparticle peaks, aided by the white dashed lines in the right panel of Fig. \ref{fig:spectrals}, showing the first-order quasiparticle peaks superimposed over the second-order spectral function. One can immediately see that in the high-density region the quasiparticle energy is shifted towards lower values by as much as by $\Delta E \sim B$. This effect becomes more substantial to the right from the angulon instabilities.In Fig. \ref{fig:peaks} we show that -- by varying the boson-boson scattering length -- the effect of the second-order correction on the position of the quasiparticle peaks becomes more conspicuous for smaller values of $a_\text{bb}$. We note that in the case of an impurity in a filled Fermi sea, a nearly perfect cancellation makes second-order corrections negligible \cite{Combescot:2008jt}.

We note that the spectral function we obtain incorrectly predicts regions with unphysical negative spectral weight, dashed in red in Fig. \ref{fig:spectrals}. These regions  always appear near the phonon branch, and originate in omission of some of the higher-order diagrams, which does not affect the quasiparticle peaks lying at lower energies. This problem has been known for a long time in the case of an electron gas \cite{Minnhagen:1974bj,Minnhagen:1975ev} and has found a solution -- applicable in all generality to any many-body diagrammatic expansion -- only recently \cite{Stefanucci:2014dk,Pavlyukh:2016bz}. In particular, it has been demonstrated that, in general, the $n$-th order truncation of a diagrammatic expansion does not lead to a positive-definite spectral function, whereas an opportune combination of all diagrams up to the $n$-th order plus certain $(n+1)$-th order diagrams is positive-definite. Following the approach in Refs. \cite{Stefanucci:2014dk,Pavlyukh:2016bz} one can divide each one of the second-order diagrams in two half-diagrams \cite{Pavlyukh:2016bz}. Then a minimal, positive-definite set of diagrams can be found by `completing the square', i.e. by introducing all third-order diagrams that can be obtained joining two half-diagrams. In the present case, one would need to include third order diagrams with the following structure:
\begin{center}
\includegraphics[width=.60\linewidth]{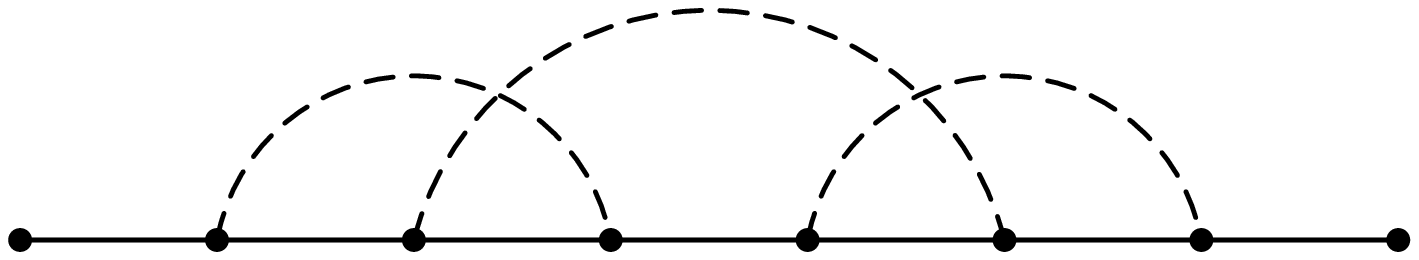} \\
\vspace{6pt} \includegraphics[width=.60\linewidth]{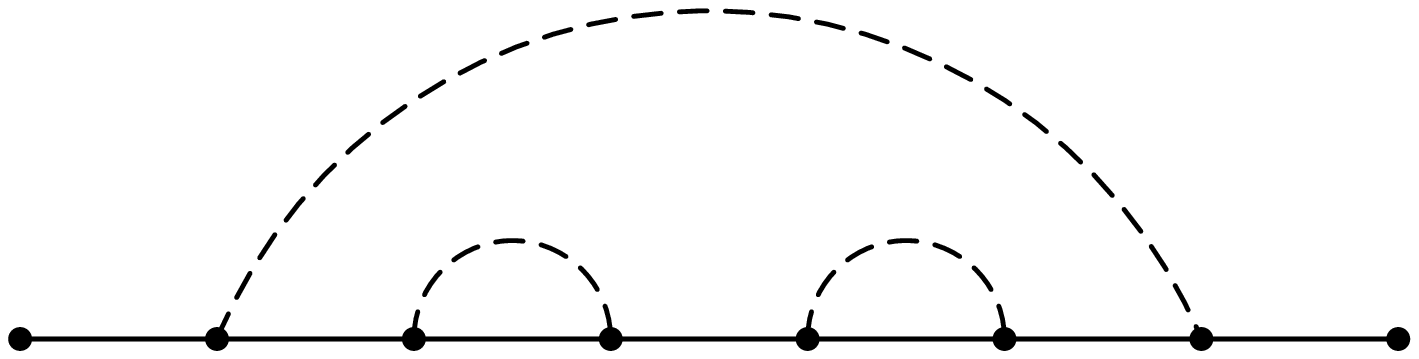} \\
\vspace{6pt} \includegraphics[width=.60\linewidth]{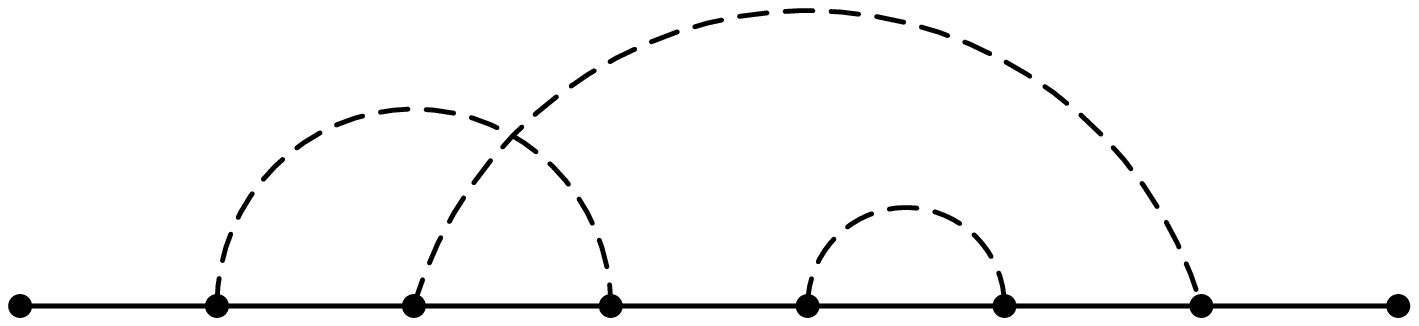} \\
\vspace{6pt} \includegraphics[width=.60\linewidth]{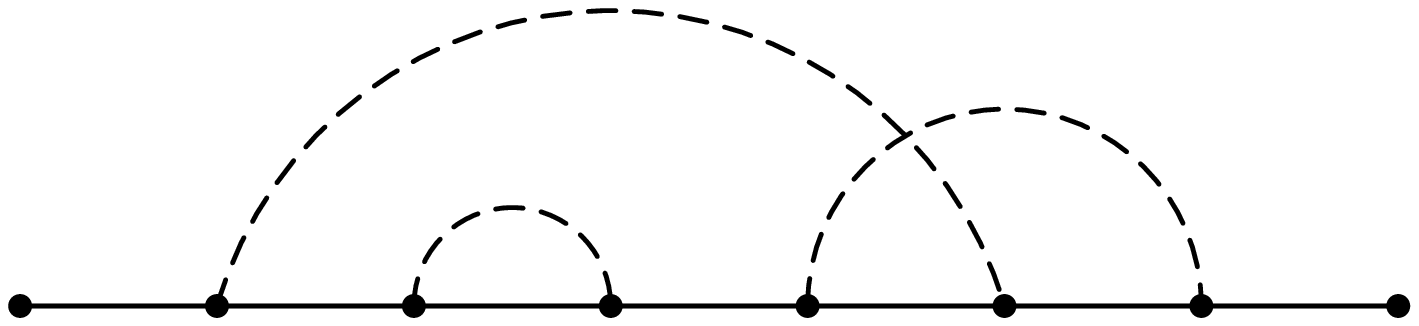}
\end{center}
whose calculation -- however made easier by the techniques introduced here -- exceeds the scope of the present work and will be the the subject of future investigations.

\subsection{Quasiparticle weight}

\begin{figure*}[t]
\centering

    \includegraphics[width=.96\linewidth]{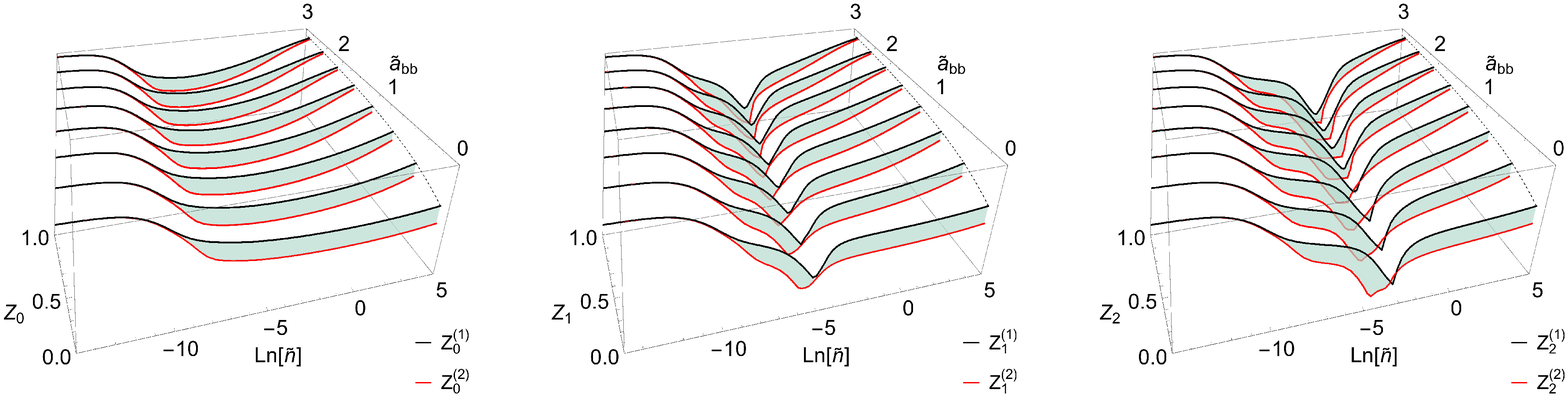}

\caption{The angulon spectral weight for the $L=0$, $L=1$ and $L=2$ states, from left to right, calculated at one-loop level (black solid line) and at two-loop level (red solid line), as a function of the dimensionless density $\tilde{n} = n (m B)^{-\nicefrac{3}{2}}$ and the dimensionless boson-boson scattering length $\tilde{a}_{\text{bb}}= a_\text{bb} (m B)^{\nicefrac{1}{2}}$.}
\label{fig:qpw}
\end{figure*}

Finally, we analyze the quasiparticle weight, $Z_L$ -- a quantity which measures the overlap between a bare particle and a dressed quasiparticle \cite{Coleman:2015}. It follows then that $Z \sim 1$ corresponds to the regime where the angulon can be accurately described as a `renormalized rotor,' whereas $Z \ll 1$ signals that the interaction with the many-body environment is hindering such a description. In the present context the quasiparticle weight is defined as~\cite{Altland:2006}:
\beq
Z^{(n)}_L=\frac{1}{1- \left. \frac{\partial \Re \Sigma_L (\omega)}{\partial \omega} \right|_{\omega = \omega_P}},
\label{eq:znl}
\eeq
where $\omega_P$ is a quasiparticle pole, corresponding to the solution of the equation
\beq
\omega_P = B L (L + 1) + \Re \Sigma_L (\omega_P),
\label{eq:pole}
\eeq
and $\Sigma_L$ is the sum of all relevant self-energy contributions, i.e.
\beq
\Sigma_L = \sum_{1 \leq j \leq n} \Sigma_L^{(j)}
\eeq
so that the superscript $n$ in Eq. (\ref{eq:znl}) refers to the order of the diagrammatic expansion. In order to understand the extent to which the two-phonon processes included in $\Sigma^{(2)}$ are affecting the properties of the angulon, in Fig.~\ref{fig:qpw} we compare the quasiparticle weights, $Z^{(1)}_L$, calculated using the one-loop theory (black solid lines) and quasiparticle weights, $Z^{(2)}_L$, calculated including both one- and two-loop contributions (red solid lines). The states with $L=0$, $L=1$ and $L=2$ are shown from left to right, as a function of the dimensionless density $\tilde{n}$ and of the dimensionless boson-boson scattering length $\tilde{a}_{\text{bb}}= a_\text{bb} (m B)^{\nicefrac{1}{2}}$.

We observe that, in general, the quasiparticle weight is close to one both for low and high values of the density, and exhibits a minimum in the intermediate density region. This picture is in agreement with the spectral function, showing angulon-phonon hybridization at work primarily for intermediate values of the density. One can see that inclusion of second-order processes by means of the two-loop diagrams results in an enhanced transfer of spectral weight from the impurity to the phonons, significantly reducing the quasiparticle weight. In the $L=0$ case, the reduction can amount to as much as $30\%$ in the parameter region we consider. For the angulon states characterised by a higher $L$, such a reduction is generally more pronounced. Furthermore, we note that the minimum of the quasiparticle weight, identifying the instability region, is shifted towards lower densities when second-order processes are taken into account. This effect arises as a result of the interplay between the angulon and phonon branches and is also evident from the spectral functions of Fig.~\ref{fig:spectrals}.

Finally, we observe that Fig. \ref{fig:qpw} shows that an increase in the boson-boson scattering length $a_\text{bb}$ stabilizes the quasiparticle description of the angulon in the density region we consider, whereas lower values of $a_\text{bb}$ correspond to a less stable angulon characterized by lower quasiparticle weights. This phenomenon is particularly evident in the high density region of each plot of Fig. \ref{fig:qpw}, where a dashed line serves as an eye-guide. This effect is somewhat reminiscent of the Landau stability criterion for the Bose gas, where the critical velocity of the particle increases with the speed of sound in the superfluid, $v_s \sim a_\text{bb}^{1/2}$~\cite{Pitaevskii:2003,Navez:1999gj}.

\section{Conclusions}

In the present paper we have introduced a path-integral treatment for the angulon. After integrating out the degrees of freedom pertaining to the many-body environment exactly, we used a perturbative treatment of the effective action  to perform a diagrammatic expansion. The resulting Feynman rules for the angulon were used to calculate the self-energy at the first and second order of the perturbative expansion. The formalism derived in this paper establishes a connection between  the theory of orbital quantum impurities -- or angulons -- and the graphical theory of angular momentum commonly used in atomic structure calculations~\cite{Balcar:2009,Judd:1963,Rudzikas:2007, Varshalovich:1988}. We exemplified the technique by revealing the role played by two-phonon processes in the angulon model in the high-density regime.

The approach we introduced significantly simplifies the treatment of orbital quantum impurities and could be naturally extended to account for more involved physical settings, e.g.\ the interaction of two angulons~\cite{Alexandrov:1995,Devreese:2009hz,Kashirina:2010eu}, or the interaction of an angulon with an external field~\cite{Redchenko:2016dq,Yakaboylu16, Rzadkowski:2017}, thereby advancing the comprehension of the angular momentum properties of quantum many-body systems. In addition, the present description of the angulon -- revolving around the angulon Green function and providing a framework for its calculation at higher orders -- paves the way to analyse  the dynamical properties of an orbital impurity. There, the diagrammatic technique is expected to be more accurate compared to the approach based on the Suzuki-Trotter decomposition of the time evolution operator~\cite{Shepperson:2017tr}.

Finally, we stress that the approach we have introduced can be incorporated into more advanced techniques that can be developed for the angulon problem, in particular those involving the analytic inclusion of higher-order terms {\cite{Stefanucci:2014dk}} or numerical calculations based on diagrammatic Monte Carlo techniques~\cite{Prokofev:1996,Prokofev:1998gz,Prokofev:1998}. The latter  represent a natural step forward for the present theory, since the diagrammatic expansions for the polaron and for the angulon feature a similar structure{~\cite{Smondyrev:1986cb,Prokofev:1998,Mishchenko:2000co}}.

\section{Acknowledgements}

We  thank Wim Casteels, Bikashkali Midya, Wojciech Rz\k{a}dkowski, Richard Schmidt, and Enderalp Yakaboylu, for valuable comments and suggestions at various stages of this work. This work was supported by the Austrian Science Fund (FWF), project Nr. P29902-N27. 

\appendix

\section{Angular momentum representation of rotationally-invariant functions of two angles}
\label{app:lemma2}

Let us consider a function of two angles, $f(\Omega,\Omega')$, which depends only on the relative angle, $\gamma$, i.e. $f(\Omega,\Omega')=f(\gamma(\Omega,\Omega'))$. Due to rotational invariance, each Green function considered in the present paper possesses this property. Then the spherical harmonics expansion $f_{l m l' m'}$, defined by
\beq
f_{l m l' m'} =  \int \mathrm{d}  \Omega \int \mathrm{d} \Omega' \ Y^*_{lm} (\Omega) Y_{l'm'} (\Omega') f(\gamma(\Omega,\Omega')),
\eeq
has the following structure:
\beq
f_{l m l' m'} = f_l \delta_{l l'} \delta_{m,m'},
\label{eq:qed}
\eeq
where the functions $f_l$ are to be defined below. Let us demonstrate it by a direct calculation. We start by rotating the spherical harmonics $Y_{l'm'} (\Omega ')$  by the angles $\left(0, - \theta, - \phi \right)$. The spherical harmonics can be expressed in the rotated frame making use of the Wigner $D$-matrix, as follows~\cite{Varshalovich:1988}
\beq
Y_{l'm'} (\Omega ') = \sum_{m''} D^{l'}_{m'' m'} (0,-\theta,-\phi) Y_{l' m''} (\Omega'')
\eeq
so that, after rearranging, we get
\begin{multline}
f_{l m l' m'} = \sum_{m''}  \int \mathrm{d} \Omega \ Y^{*}_{lm} (\Omega) D^{l'}_{m'' m'} (0,-\theta,-\phi) \times \\ \times \int \mathrm{d} \Omega'' Y_{l' m''} (\Omega'') f(\gamma(0,\Omega''))
\label{eq:flx}
\end{multline}
where (with a slight abuse of notation) $\gamma(0,\Omega'')$ is the angle between the north pole and the point on the surface of a sphere identified by the angle $\Omega''$, and clearly $\gamma(\Omega,\Omega') = \gamma(0,\Omega'')$. The innermost integral can be readily evaluated using the standard representation of spherical harmonics in terms of the Legendre polynomials $P_l$, giving
\beq
\int \mathrm{d} \Omega'' \ Y_{l' m''} (\Omega'') f(\gamma(0,\Omega'')) = f_{l'} \sqrt{\frac{2 l' + 1}{4 \pi}} \delta_{m'' 0}
\label{eq:legendre}
\eeq
where
\beq
f_l = 2 \pi \int_{-1}^1 \mathrm{d} x \ P_l (x) f(x),
\label{eq:flxx}
\eeq
with a substitution of $x=\cos \gamma$. Eq.~(\ref{eq:flxx}) defines the expansion of a rotationally-invariant function of two angles in the angular momentum basis, and will be used throughout the  paper. With this definition, equation (\ref{eq:flx}) becomes
\beq
f_{l m l' m'} = f_{l'} \sqrt{\frac{2 l' + 1}{4 \pi}} \int \mathrm{d} \Omega \ Y^{*}_{lm} (\Omega) D^{l'}_{0 m'} (0,-\theta,-\phi),
\eeq
from which, using the relation \cite{Varshalovich:1988}
\beq
D^{l}_{0 m} (-\chi,-\theta,-\phi) = \sqrt{\frac{4 \pi}{2l + 1}} Y_{l m} (\theta,\phi),
\eeq
along with the orthogonality and symmetry properties for spherical harmonics, we recover Eq. (\ref{eq:qed}).

\section{Green functions in the angular momentum basis}
\label{app:gf}

The Green function for a free quantum rotor can be written as~\cite{Dickhoff:2005,Favro:1960ie}
\beq
G_0(\Omega,\Omega'; t )= - \mathrm{i} \sum_n \psi_n(\Omega) \psi_n^*(\Omega') e^{-\mathrm{i} E_n t},
\label{eq:appg0begin}
\eeq
where the index $n$ runs over all the eigenstates $E_n$ of the rotor, each one corresponding to a wavefunction $\psi_n$. Before taking the Fourier transform, we ensure causality by inserting a step function, which corresponds to calculating a retarded propagator. We use the following integral representation for the step function
\beq
\theta (t) = - \int \frac{\mathrm{d} E'}{2 \pi \mathrm{i}} \frac{e^{-\mathrm{i} E' t}}{E' + \mathrm{i} \delta}
\eeq
where the limit $\delta \to 0^+$ is implied. The retarded Green function in frequency representation is then given by
\beq
G_0(\Omega,\Omega'; \omega) = \int_{-\infty}^{+\infty} \mathrm{d} t \ e^{\mathrm{i} \omega t} \ G_0(\Omega,\Omega'; t ) \theta(t)
\eeq
After carrying out the integrations we get the Lehmann spectral representation for the retarded Green function
\beq
G_0(\Omega,\Omega'; \omega) = \sum_n \frac{\psi_n(\Omega) \psi_n^*(\Omega')}{\omega - E_n + \mathrm{i} \delta} \; .
\eeq

In the case of a linear rotor, the wavefunctions $\psi_n$ are given by the spherical harmonics with $n=\{ \lambda, \mu\}$, so that we get:
\beq
G_0 (\Omega,\Omega' ; \omega) = \sum_{\lambda \mu} \frac{Y^{*}_{\lambda \mu} (\Omega) Y_{\lambda \mu} (\Omega')}{\omega - B \lambda (\lambda + 1) + \mathrm{i} \delta} \; .
\eeq
The sum over $\mu$ can be carried out using the spherical harmonics addition theorem, obtaining
\beq
G_0 (\Omega,\Omega' ; \omega) = \sum_\lambda \frac{2 \lambda + 1}{4 \pi} \frac{P_\lambda (\cos \gamma(\Omega,\Omega'))}{\omega - B \lambda (\lambda + 1) + \mathrm{i} \delta}
\eeq
where $\gamma(\Omega,\Omega')$ is the angle between $\Omega$ and $\Omega'$. Writing the result in the angular momentum basis as outlined in Appendix \ref{app:lemma2}, and using the orthogonality of Legendre polynomials, we obtain:
\beq
G_{0,\lambda} (\omega) = \frac{1}{\omega - B \lambda (\lambda + 1) + \mathrm{i} \delta}
\label{eq:g0end}
\eeq
Likewise, we can derive the interaction propagator in the angular momentum basis, starting from its definition
\beq
\chi (\Omega,\Omega' ; t) = -\mathrm{i} \sum_\lambda P_\lambda (\cos \gamma(\Omega,\Omega')) \mathcal{M} (t) \; .
\label{eq:chi0}
\eeq
The calculation for the free case leading from Eq. (\ref{eq:g0begin}) to Eq. (\ref{eq:g0end}) is straightforward to adapt to the present case. In a completely analogous way, taking the Fourier transform and using the spherical harmonics expansion, we obtain
\beq
\chi_\lambda (\omega) = \sum_k \frac{|U_\lambda(k)|^2}{\omega - \omega_k + \mathrm{i \delta}} \; .
\label{eq:chifinal}
\eeq

\section{The convolution theorem in the angular momentum basis}
\label{app:convolution}

The convolution theorem states that the Fourier transform of a convolution is the product of the Fourier transforms of every single function entering the convolution. We would like to find an analogous results for the spherical harmonics expansion, holding between the spherical basis and the angular momentum basis. In order to do so we consider the following `spherical convolution':
\beq
h(\Omega_i, \Omega_f) = \int \mathrm{d} \Omega' f(\Omega_i,\Omega') g(\Omega',\Omega_f) \;.
\eeq
Again, and crucially, we assume that $f$ and $g$ depend only on the angle between their arguments, i.e.
\beq
f(\Omega_1,\Omega_2) = f(\gamma(\Omega_1,\Omega_2)),
\eeq
and similarly for $g$. Rotational invariance implies that $h$ should be a function only of the angle between its arguments as well, so we can write without loss of generality:
\beq
h(\Omega_i,\Omega_f) = h(\gamma(\Omega_i,\Omega_f)) \; .
\eeq
We start from expanding $h$ in the angular momentum basis, making use of Eq. (\ref{eq:legendre})
\beq
h_{l} = \sqrt{\frac{4 \pi}{2l + 1}} \int \mathrm{d} \Omega \ Y_{l0} (\Omega) \int \mathrm{d} \Omega' \ f(\gamma(0,\Omega')) g(\gamma(\Omega',\Omega)) 
\eeq
Following the analogy with  Appendix \ref{app:lemma2}, by inverting the integration order and rotating the spherical harmonics by the angle $(0,-\theta,-\phi)$, we obtain
\begin{multline}
h_{l} =  \sqrt{\frac{4 \pi}{2l + 1}}  \sum_{m} \int \mathrm{d} \Omega' f(\gamma(0,\Omega')) D^l_{m 0} (0,-\theta',-\phi') \times \\ \times \int \mathrm{d} \Omega'' \ Y_{lm} (\Omega'') g(\gamma(\Omega'',0))
\end{multline}
where the integrals appear decoupled. Making use of Eq.~(\ref{eq:legendre}), the innermost integral is easily seen to be
\beq
g_l \sqrt{\frac{2 l + 1}{4 \pi}} \delta_{m 0}
\eeq
After carrying out the summation over $m$, the innermost integral can be evaluated using, again, the techniques in Appendix \ref{app:lemma2}, which gives $f_l$. Combining the results we finally get
\beq
h_l = f_l \ g_l
\label{eq:hlmfinal}
\eeq

which extends the usual convolution theorem for the Fourier transform to the case of the spherical harmonics expansion. The result just found can be readily extended to the `spherical convolution' of an arbitrary number of rotationally invariant functions. Introducing the $\star$ notation for the convolution in the spherical basis, Eq. (\ref{eq:hlmfinal}) takes the form

\beq
(f \star g)_{l} = f_l \ g_l
\eeq

Extending by induction, the convolution theorem for $n$ functions reads
\beq
(f_1 \star f_2 \star \ldots \star f_n)_{l} = \prod_{j=1}^n (f_j)_l  \; .
\eeq

\section{Graphical representation of the geometric terms}
\label{app:graphical}

Let us focus on the resemblance between the rules listed in Table \ref{table1} and the rules derived within the graphical theory of angular momentum. Such a similarity paves the way to develop a formal connection with the diagrammatic theory presented in this paper and allowing for a great simplification of otherwise cumbersome calculations. Let us take into account the `geometric' contributions to the self-energy, as defined in Section \ref{sec:sigmaaz}, i.e.
\begin{widetext}
\beq
\Sigma_\lambda^{(2,A) \text{geom}} = \sum_{\{\mu_i\}} (-1)^{\sum_i \mu_i} (-1)^{\lambda_1 + \lambda_2 + \lambda_3 + \lambda} \begin{pmatrix} \lambda & \lambda_1 & \lambda_2 \\
\mu & -\mu_1 & -\mu_2
\end{pmatrix} \begin{pmatrix} \lambda_2 & \lambda_3 & \lambda_4 \\
\mu_2 & -\mu_3 & -\mu_4
\end{pmatrix} \begin{pmatrix} \lambda_1 & \lambda_4 & \lambda_5 \\
\mu_1 & \mu_4 & -\mu_5
\end{pmatrix} \begin{pmatrix} \lambda_3 & \lambda_5 & \lambda \\
\mu_3 & \mu_5 & -\mu
\end{pmatrix}
\eeq
and
\beq
\Sigma_\lambda^{(2,B) \text{geom}} = \sum_{\{\mu_i\}} (-1)^{\sum_i \mu_i} (-1)^{\lambda_1 + \lambda_2 + \lambda_3 + \lambda} \begin{pmatrix} \lambda & \lambda_1 & \lambda_2 \\
\mu & -\mu_1 & -\mu_2
\end{pmatrix} \begin{pmatrix} \lambda_2 & \lambda_3 & \lambda_4 \\
\mu_2 & -\mu_3 & -\mu_4
\end{pmatrix} \begin{pmatrix} \lambda_3 & \lambda_4 & \lambda_5 \\
\mu_3 & \mu_4 & -\mu_5
\end{pmatrix} \begin{pmatrix} \lambda_1 & \lambda_5 & \lambda \\
\mu_1 & \mu_5 & -\mu
\end{pmatrix}
\eeq
\end{widetext}

\begin{figure}[ht!]
\centering

    \includegraphics[width=.75\linewidth]{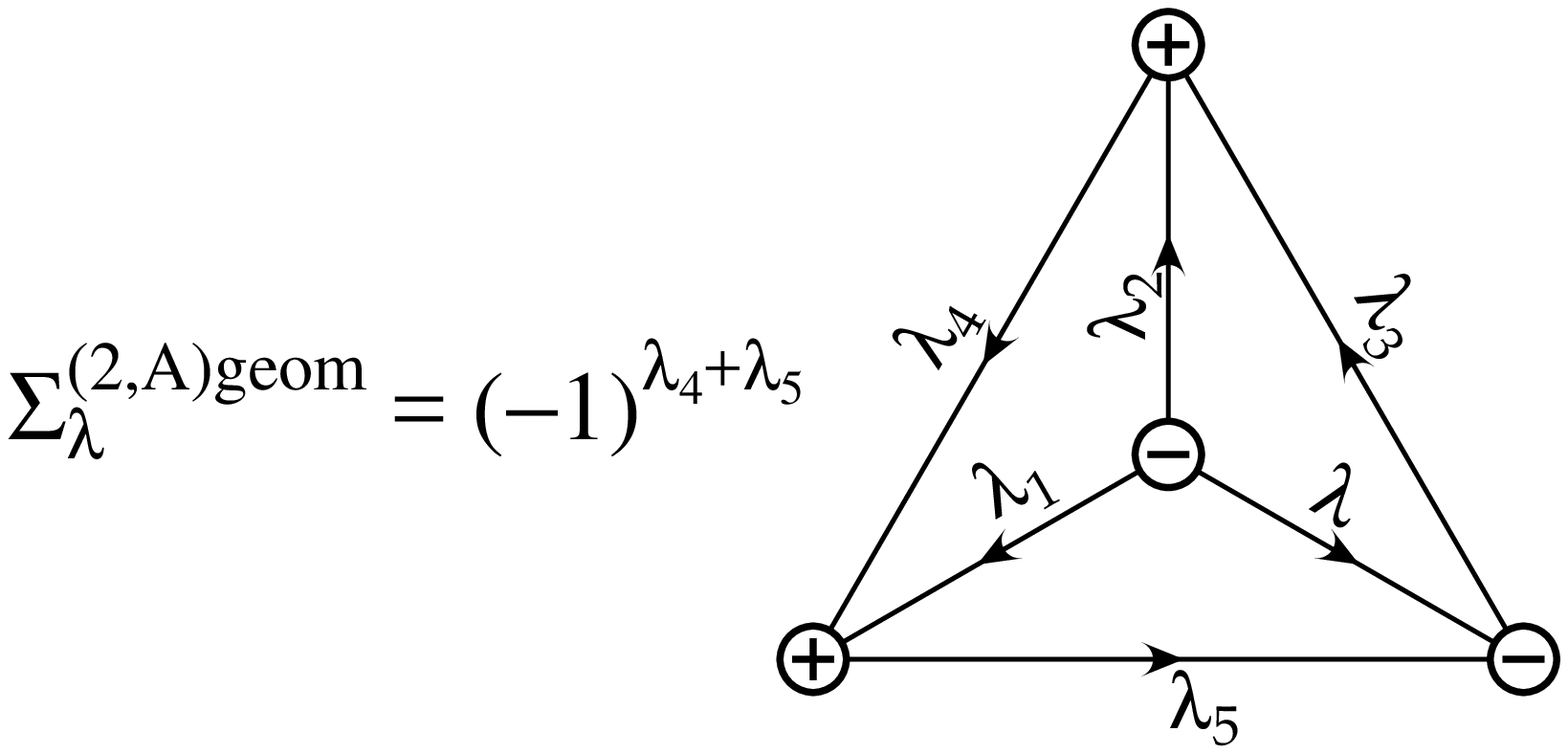}\\[10pt]
    \includegraphics[width=.75\linewidth]{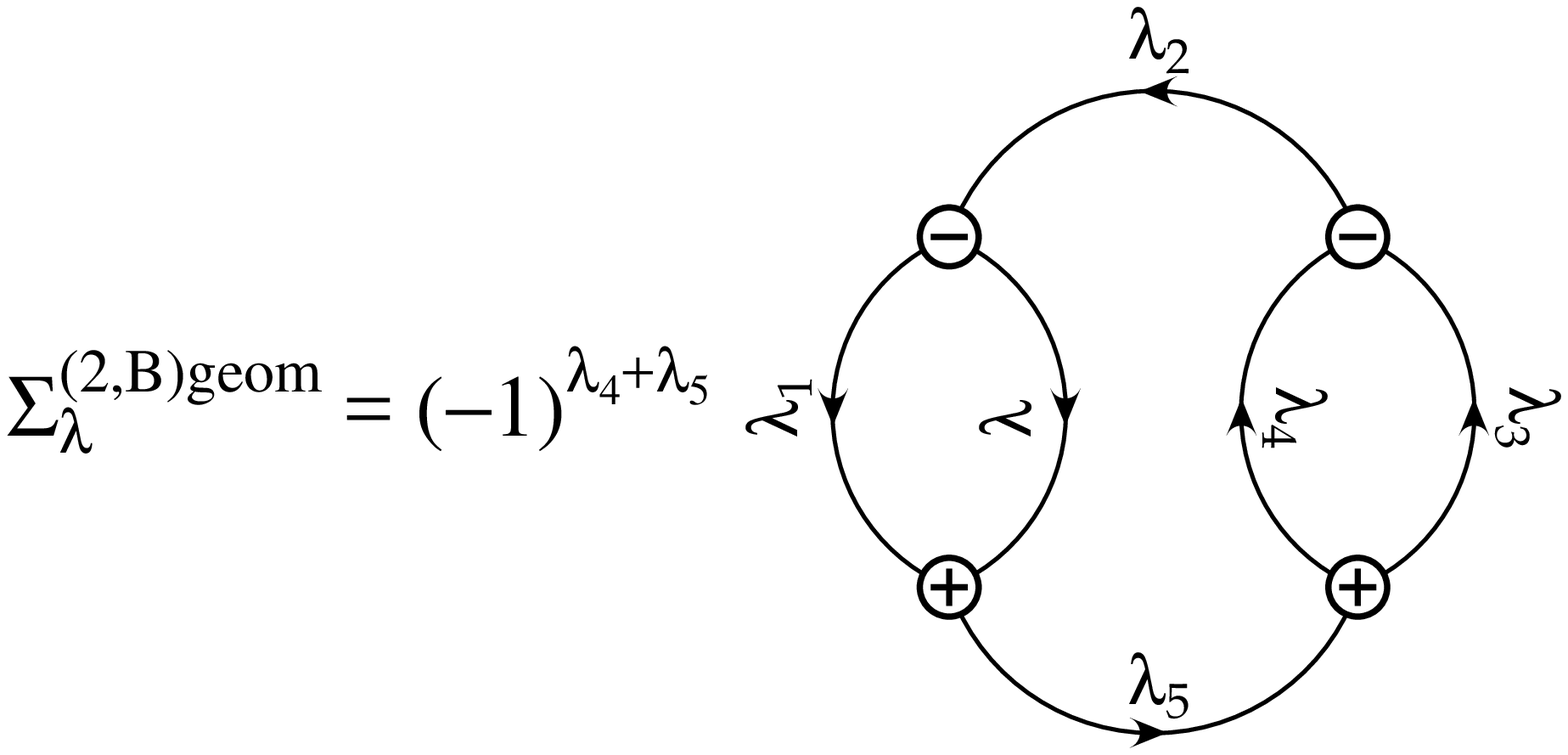}

\caption{The diagrams in Fig. \ref{fig:second} and the corresponding analytic expressions for the geometric part can be readily mapped onto diagrams of the graphical theory of angular momentum~\cite{Balcar:2009}. These diagrams encode the geometric aspect of the self-energies and reflect  the conservation of angular momentum. These diagrams still contain the summations over $\mu_i$ which can be eliminated using graphical techniques: the upper diagram is simplified by changing the sign of two nodes, thus reconstructing the $6j$ symbol, whereas the lower diagram is simplified making use of the separation technique for subdiagrams connected by two lines. As a result we arrive at the noticeably simpler results of Eq. (\ref{eq:s2ageoapp}) and Eq. (\ref{eq:s2bgeoapp}).}
\label{fig:gtam}
\end{figure}

We rewrite these analytic expressions using the rules of the graphical theory of angular momentum. In particular, we adopt  the conventions of Ref. \onlinecite{Balcar:2009}. As already noted, the resulting diagrams have exactly the same topological structure as their `parent' diagrams shown in Fig. \ref{fig:second}, provided that the lines corresponding to the initial and final states are joined and every interaction line is substituted with a solid line. After elementary manipulations, these diagrams can be converted into the diagrams shown in Fig. \ref{fig:gtam}. Finally, using the rules of the graphical theory of angular momentum \cite{Balcar:2009} we can carry out the summations over $\mu_i$. In particular, in the case of $\Sigma_{\lambda}^{(2,A)\text{geom}}$ we can modify the orientation of two lines and change the sign of the negative nodes, obtaining the representation of the $6j$ symbol. As a result, we obtain Eq. (\ref{eq:s2ageo}) which we present here for the sake of completeness:

\beq
\Sigma_\lambda^{(2,A) \text{geom}} = \frac{(-1)^{\lambda_1 + \lambda_2 + \lambda_3 + \lambda_4}}{2 \lambda + 1} \begin{Bmatrix}
\lambda_2 & \lambda_1 & \lambda \\
\lambda_5 & \lambda_3 & \lambda_4
\end{Bmatrix}
\label{eq:s2ageoapp}
\eeq
Similarly, in the case of $\Sigma_{\lambda}^{(2,B)\text{geom}}$, we need to apply the graphical rules for the separation of two internal lines, on the $\lambda_2$ and $\lambda_5$ lines. The resulting simplified graphical representation leads to Eq. (\ref{eq:s2bgeo}):

\beq
\Sigma_\lambda^{(2,B) \text{geom}}
 = \frac{(-1)^{\lambda_4 + \lambda_5} \delta_{\lambda_2,\lambda_5}}{(2 \lambda + 1)(2 \lambda_5 + 1)} \{\lambda~\lambda_1~\lambda_2\} \{\lambda_2~\lambda_3~\lambda_4\} \; .
 \label{eq:s2bgeoapp}
\eeq
Although exemplified with one- and two-loops diagrams, as for most  results of the present paper this connection is valid at every order in the diagrammatic expansion.

\FloatBarrier
\bibliography{references,xtra}

\end{document}